\documentclass[fleqn,usenatbib]{mnras}

\usepackage{newtxtext,newtxmath}

\usepackage[T1]{fontenc}

\DeclareRobustCommand{\VAN}[3]{#2}
\let\VANthebibliography\thebibliography
\def\thebibliography{\DeclareRobustCommand{\VAN}[3]{##3}\VANthebibliography}

\usepackage{graphicx}	
\usepackage{amsmath}	

\usepackage{times}
\usepackage{float}
\usepackage{hyperref}
\usepackage{scrlayer-scrpage}
\usepackage{xcolor}
\definecolor{orange}{rgb}{0.93, 0.57, 0.13}
\usepackage{scrlayer-scrpage}
\usepackage{caption}
\usepackage{subcaption}
\usepackage{hyperref}
\usepackage{colortbl}
\usepackage{float}
\usepackage{fix-cm}

\title[Short title, max. 45 characters]{Feedback and Star Formation Efficiency in High-Mass Star-Forming Regions}

\author[B. Zimmermann et al.]{
Birka Zimmermann$^{1}$\thanks{E-mail: zimmermann@ph1.uni-koeln.de},
Stefanie Walch$^{1,2}$,
Seamus D. Clarke$^{3,4}$,
Richard W\"unsch$^{5}$,
Andre Klepitko$^{1}$
\\
$^{1}$ I. Physikalisches Institut, Universität zu Köln, Zülpicher Str. 77, D-50937 Köln, Germany\\
$^{2}$Center for data and simulation science, University of Cologne, www.cds.uni-koeln.de\\
$^{3}$Department of Physics, National Cheng Kung University, 70101 Tainan, Taiwan\\
$^{4}$Institute of Astronomy and Astrophysics, Academia Sinica, No. 1, Section 4, Roosevelt Road, Taipei 10617, Taiwan\\
$^{5}$Astronomical Institute of the Czech Academy of Sciences, Bo\v{c}n\'{i} II 1401, 141 00 Prague, Czech Republic}

\date{Accepted XXX. Received YYY; in original form ZZZ}

\pubyear{2025}

\begin{document}
\label{firstpage}
\pagerange{\pageref{firstpage}--\pageref{lastpage}}
\maketitle

\begin{abstract}
To advance our understanding of massive star formation, it is essential to perform a comprehensive suite of simulations that explore the relevant parameter space and include enough physics to enable a comparison with observational data. 
We simulate the gravitational collapse of isolated, parsec-scale turbulent cores using the FLASH code, modelling stars as sink particles. Our simulations incorporate ionizing radiation and the associated radiation pressure from stellar sources, and non-ionizing radiation and its dust heating, along with self-consistent chemistry, to capture the properties of emerging ultra-compact HII regions. Dust, gas, and radiation temperature are computed independently. The initial conditions are informed by ALMAGAL observations. 
We assess stellar feedback, comparing ionizing radiation and radiation pressure. Ionizing radiation ultimately halts mass accretion on to sink particles, while direct radiation pressure enhances the expansion of HII regions. Heating from non-ionizing radiation suppresses fragmentation. 
We examine the effect of spatial resolution, finding that higher resolution leads to more sink particles which are situated in environments with higher densities. As a result, ionizing radiation remains trapped longer, allowing continued accretion and yielding a higher overall star formation efficiency (SFE). 
We explore the impact of varying initial conditions, including the core density profile, virial parameter, and metallicity. Our parameter study reveals that a flatter density profile, higher virial parameter, and increased metallicity promote fragmentation, potentially enhancing the SFE by slowing the growth of the most massive stars and delaying the onset of stellar feedback. Overall, we find SFEs between 35\% and 57\%. Stellar feedback dictates the final SFE.

\end{abstract}




\begin{figure*}
	\includegraphics[width=1.0\textwidth,trim=0.0cm 9.0cm .0cm 12.0cm,clip]{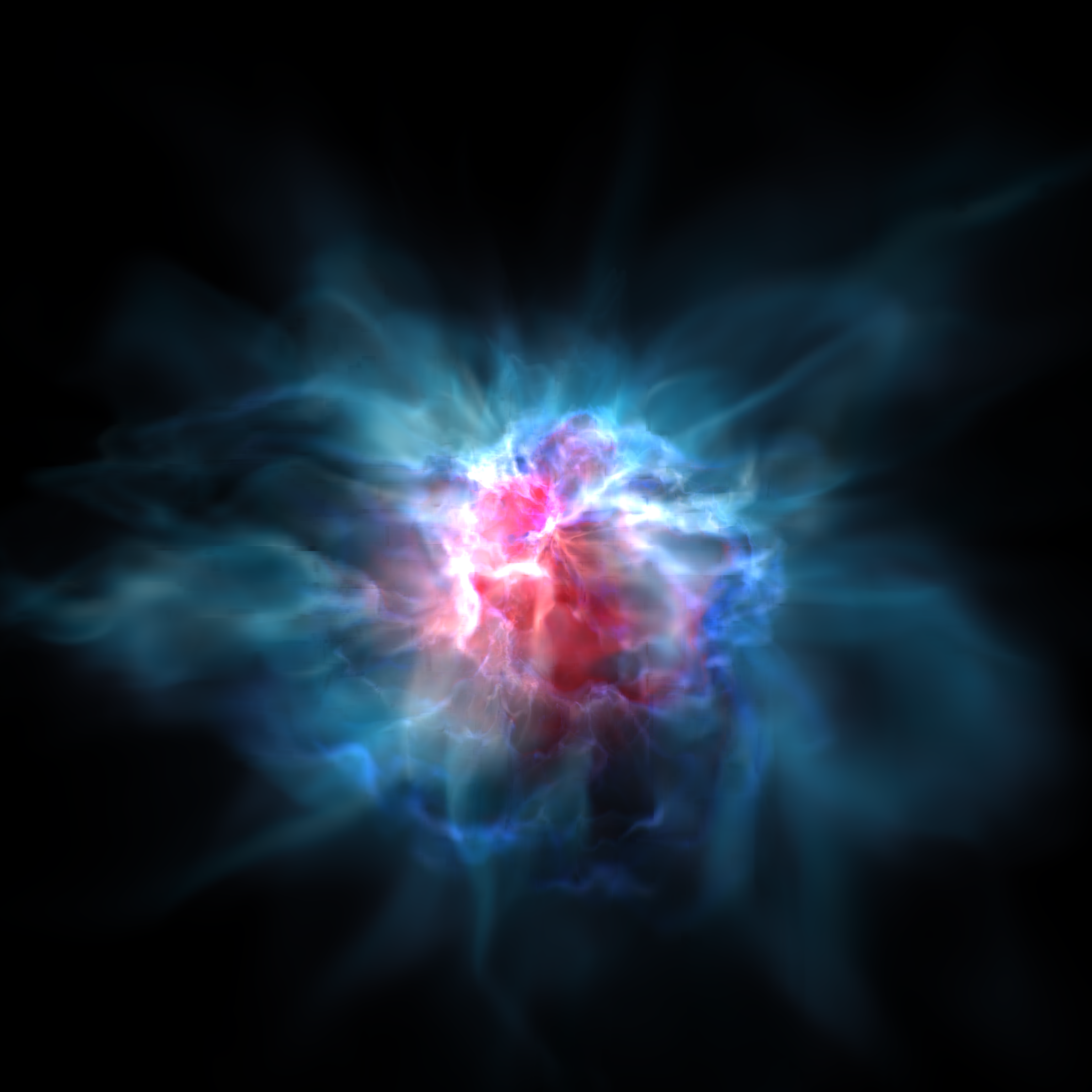}
    \caption{Looking into the heart of a massive star-forming core. Volume rendering of our run \textsc{RFL10} shown at $0.78 \, t_{\rm ff}$. 
    We render molecular hydrogen, H$_2$, showing the cold gas distribution within the core. In addition, we show atomic hydrogen, H, and ionized hydrogen, H$^{+}$ in blue and red, respectively. The alpha channels are proportional to the logarithmic density.
    The volume rendering reveals a violent outflow accompanied by an expanding ultra-compact HII region with an asymmetric shape, supported by radiation pressure. The atomic gas (blue) surrounds the ionized gas (red), which fills the bubble.}
    \label{fig:volren}
\end{figure*}

\section{Introduction}
Star formation is a fundamental process, shaping both the large and small astronomical scales and simultaneously influencing galactic dynamics \citep{2013Bolatto.499..450B}. One of the largest unresolved problems in modern star formation is the formation of massive stars (here defined as $>$8 M$_\odot$). The formation of low-mass stars ($<$8 M$_\odot$) is mainly governed by the interplay of gravity, turbulence, and thermal pressure leading to quasi-Jeans fragmentation; however, additional support against gravity is necessary for the formation of higher-mass stars \citep{mckeetan2003}. Such support is thought to be provided primarily by magnetic fields \citep{commerccon2011collapse,Commercon2022,mignon-risse2021,Hennebelle2019} and radiative feedback \citep{Krumholz_2009}. It is a fact that high-mass stars evolve faster and start nuclear burning before the mass accretion process is finished, so feedback and accretion happen at the same time \citep{2009Hosokawa}. Cores hosting massive star formation are very dense and have higher average temperatures ($\sim$20~K) than lower mass and less dense cores ($\sim$10~K) \citep{Beuther2007}. 
A manifest of young massive stars is the presence of ultra-compact HII (UC-HII) regions which have sizes of $\sim 0.1 \, \mathrm{pc}$ and densities of around $10^{4} \, \mathrm{cm}^{-3}$ \citep[][]{Wood,churchwell2002ultra, hoare2006ultracompact}.

The formation of high-mass stars is associated with very energetic feedback ((ionzing) radiation, radiation pressure, protostellar outflows and stellar winds). At the end of their lifespans supernovae feedback becomes important and influences the interstellar medium (ISM) on larger scales. Several numerical simulations have been performed to model feedback in the ISM e.g., the \textsc{SILCC} project \citep{walch2015,Peters2017,Rathjen2021,Rathjen2023}, the \textsc{TIGRESS} simulations \citep{2018KimOstriker,KimKim2023} or the \textsc{STARFORGE} project \citep{2021Gudric,Guszejnov2022}.

However, the feedback from massive stars will only affect their environment on parsec scales or beyond once they have dispersed their parental molecular cloud core.
Massive stars are initially deeply embedded, causing a delay between the birth of a massive star and the time from which it blows a larger-scale bubble.
In this initial phase, radiation pressure (RP) on gas and dust becomes important \citep{Rosen2019,2023Kleptiko,2021Ali}.
\cite{2018Kuiper} find that RP may reduce the mass accretion rate while stars are still forming and thus limits their final mass.
Not only RP, but also radiative heating due to the absorption of ionizing ($> 13.6 \, \rm eV$) and non-ionizing ($<13.6 \, \rm eV$) stellar radiation by the surrounding gas and dust is important in this phase \citep{Rosdahl2013,2023Kleptiko}. Before an UC-HII region emerges, large scale protostellar outflows are commonly observed \citep{2021Avison}. Stellar isotropic winds and collimated protostellar outflows are assumed to reduce the mass accretion rate of the forming massive stars \citep{Kuiper2015,Oliva2023, 2022Rosen}. However, these processes may not represent the ultimate mechanism responsible for halting accretion \citep{2016Kuiper, Rosen2020, 2022Rosen}.
The interplay of the different feedback processes plays an important role in setting the star formation efficiency in massive host cores, and there are still open questions regarding how they act in concert.

Moreover, the impact of the initial conditions on the formation process of massive stars as well as their mass evolution and spatial distribution is not well understood \citep[e.g.,][]{2014Tan}.  In particular, key parameters such as density profile, virial parameter, and metallicity are expected to play crucial roles.

\textsc{Density Profile.}
The evolution of the cloud core strongly depends on the initial density profile \citep{Boss1987,Lomax2013}. Existing numerical simulations commonly use density profiles that vary from a flat distribution to a very steep power-law profile. Using a core mass of $100 \, M_\odot$, \cite{Girichidis2011} compared four different density profiles (uniform,
Bonnor-Ebert type and power law density profiles) with $\rho(r)\sim r^{-1.5}$ and $\rho(r) \sim r^{-2}$, while neglecting any type of radiation. They find that a steep density profile shows a rapid collapse and leads to a central high-mass star, while a flat core has a longer collapse phase and several lower-mass stars are produced. In general, low-mass stars form later and also via disc fragmentation \citep{2007Stamatellos}, which requires sufficient resolution to capture. However, in the case of massive stars, disc fragmentation is a complex process that may also depend sensitively on radiative feedback, as different treatments of radiation can significantly influence disc stability and fragmentation behaviour \citep[e.g.,][]{2016rosen,2020Oliva,2010Kuiper}. Additionally, magnetic fields, which are not included in this study, can provide additional support against gravitational collapse and reduce fragmentation by stabilizing discs and regulating angular momentum transport \citep{Rosen2020}.

\textsc{Turbulence/ Virial Parameter.}
The initial turbulent velocity field dictates the morphology of the gas and dust during the initial core collapse \citep{Walch2010}. Especially the virial parameter is important for the fragmentation process \citep{Girichidis2011}. 

\textsc{Metallicity.}

The metallicity in the natal star-forming environment impacts the thermal evolution of prestellar cores \citep{Omukai2005}.
Metal lines provide efficient cooling, particularly in the hotter, lower-density regions influenced by stellar feedback \citep{Schneider2003}.
However, at the higher densities typical of prestellar cores, dust cooling dominates and plays a crucial role in regulating the temperature. Therefore, decreasing metallicity generally implies a lower dust content, leading to less efficient dust cooling and consequently higher gas and dust temperatures. This increase in temperature raises the Jeans mass, reducing fragmentation, especially the formation of lower-mass stars \citep{2022Chon,2023Kleptiko}. In the extreme case of Population III star formation, this process can result in a top-heavy stellar initial mass function (IMF) \citep[e.g.,][]{2013Whalen,Latif2022}.

In this paper, we investigate the influence of ionizing radiation and the associated radiation pressure \citep{2023Kleptiko} on the massive star formation process as well as the impact of the spatial resolution. Furthermore, we carry out a parameter study of high-mass star-forming regions, where we vary the virial parameter, density profiles, and metallicities. The parameter range study is informed by the ALMAGAL\footnote{ALMA Large Program 2019.1.00195.L} observations, which is an ALMA large-scale programme and includes observations from over thousand high-mass star-forming regions \citep{2025Molinari, 2025Sanchez, 2025Coletta}. Thus, in a follow-up paper, we will confront observations and simulations. 

This paper is structured as follows. First, we explain the numerical methods and initial conditions of the simulations in section \ref{sec:SimulationDetails}, where we highlight the treatment of ionizing radiation and RP. In section \ref{sec:Analysis}, we introduce a fiducial run and show the importance of ionizing radiation and RP as well as the impact of numerical resolution on the cloud evolution. In section \ref{sec:ParameterStudy} we investigate the results of the parameter study, followed by a discussion in section \ref{sec:Discussion} and conclude in section \ref{sec:Conclusion}.
\begin{figure*}
	\includegraphics[width=1.0\textwidth,trim=0.5cm 1cm 0.5cm 1.5cm]{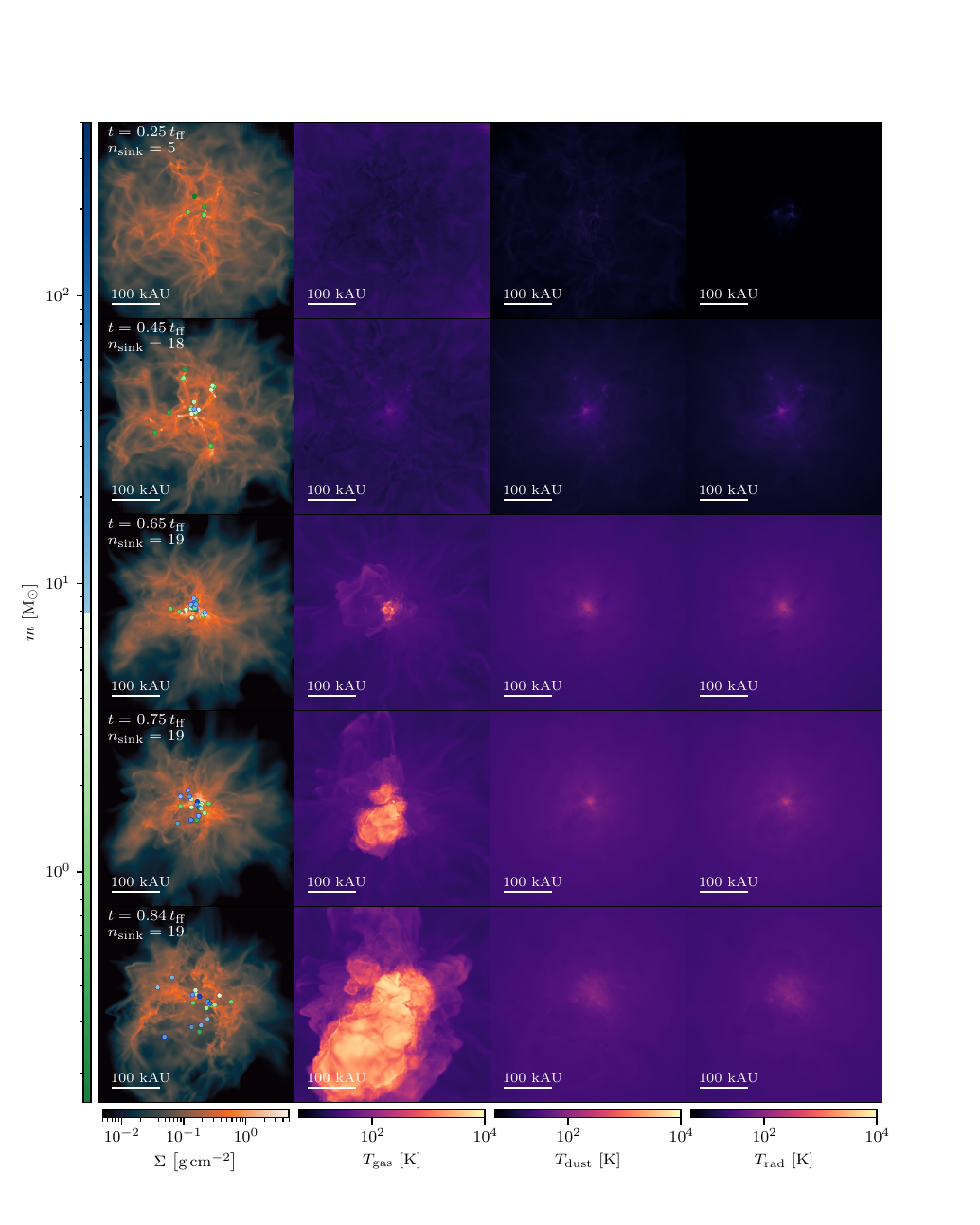}
    \caption{Time evolution of the fiducial run (from top to bottom). From left to right we show the projection in $z$-direction of the column density $\Sigma$ and the mass-weighted temperatures of gas $T_\mathrm{gas}$, dust $T_\mathrm{dust}$, and radiation $T_\mathrm{rad}$. Small circles represent sink particles. A green colour scheme represents lower-mass stars, while a blue colour scheme shows more massive sinks ($>8 \mathrm{M}_\odot$). After $ \sim 0.4 t_\mathrm{ff}$ (where $t_\mathrm{ff}=0.526$~Myr; see Table~\ref{tab_parameter}) massive sink particles are formed and drive an UC-HII region.}
    \label{fig:fiducialrun}
\end{figure*}

\section{Simulation Details}
\label{sec:SimulationDetails}
\subsection{Numerical Methods}
\label{sec:NumMeth}
We use the adaptive mesh refinement (AMR) code FLASH 4.6 \citep{fryxel2000} to calculate (magneto-)hydrodynamical prestellar core collapse simulations. 
Hydrodynamics is solved using the entropy-stable scheme by \citet{2016Derigs}.
To follow the chemical evolution of the gas, we use a non-equilibrium chemical network that tracks the evolution of seven species on the fly \citep{Glover2007,Glover2010}. These species are molecular, atomic, and ionized hydrogen as well as carbon monoxide, ionized carbon, oxygen, and free electrons (H$_2$, H, H$^+$, CO, C$^+$, O, e$^-$). The network uses reaction rates based on \cite{Nelson_1997} and our implementation of all heating and cooling processes within is summarized in \citet{walch2015}. We assume that gas and dust are perfectly mixed. 
The dust-to-gas mass ratio, $f_\mathrm{d}$, depends on the gas metallicity, $Z$, and is calculated by
\begin{eqnarray}
    f_d &=& \frac{1}{100} \frac{Z}{Z_\odot} \, .
    \label{eq:fd}
\end{eqnarray}

We use a tree-based solver for self-gravity and diffuse radiation \citep{wunsch2018}. \textsc{TreeRay/OpticalDepth} is used to calculate the dust shielding as well as the self-shielding of molecular hydrogen and CO.
Here, we neglect an external interstellar radiation field (ISRF) as the cores are highly shielded and supposedly embedded in a dense environment.
Additionally, the radiative transport module \textsc{TreeRay/OnTheSpot} deals with ionizing radiation from point sources (e.g. sink particles representing stars) in a reverse ray-tracing approach \citep{wunsch2021}. The contributions of all cells and sink particles are mapped onto rays. We use 48 rays which are distributed according to the HEALPix algorithm \citep{2005Gorski}. Along each ray, the absorption of ionizing radiation by gas and the resulting gas heating and ionization are calculated \citep{2018Haid}. For emission, we assume the On-The-Spot approximation \citep{Osterbrock_1988}. 

The module \textsc{Treeray/RadPressure} described in \cite{2023Kleptiko} is used to treat non-ionizing radiation
where every cell and sink particle is a source of non-ionizing radiation. 
The integration of each ray along its line of sight gives a radiative intensity. Taking the contribution of all integrated rays into account a radiative flux and a mean intensity is calculated.
Non-ionizing radiation is absorbed by the dust and thus heats the dust (see below). 
The dust opacity, $\kappa_\mathrm{P}$, depends on the metallicity and temperature, $T$, and is given by \citep[see][]{semenov, Krumholz2012, 2023Kleptiko}:
\begin{eqnarray}
    \kappa_\mathrm{P}(T,Z) = 10^{-1} \frac{Z}{Z_\odot} \frac{\mathrm{cm^2}}{100 \,\mathrm{K^2\, g}} \times 
    \begin{cases}
    T^2 & {\rm for}\; T<150\,\mathrm{K} \\
    {(150\,\mathrm{K})^2} & \mathrm{else}
    \end{cases}\,.
    \label{eq:dustOpacity}
\end{eqnarray}
In this set of simulations the dust is heated according to the non-ionizing radiation intensity, but the RP momentum of non-ionizing radiation is not transferred to the dust.
In contrast, \textsc{Treeray/RadPressure} calculates the RP of ionizing radiation onto gas and dust and transfers the corresponding momentum. The heating from ionizing radiation is also included and handled via the chemistry network \citep[as described in][]{2018Haid}.

The temperatures of dust, gas, and radiation are calculated self-consistently by balancing various heating and cooling processes and thus, can differ from each other (see Fig.~\ref{fig:fiducialrun}).
We include cooling due to collisional dissociation of H$_2$, collisional ionization of H and metal line cooling, in particular of C$^+$ and O \citep{walch2015}. The cooling of the gas by dust-gas coupling is also treated. Hereby, the dust temperature is calculated self-consistently, assuming it to be in thermal equilibrium. \citep[see e.g.][ their Sec. 4.6]{2023Kleptiko}. In fully ionized gas within the forming HII regions, cooling by H$^+$  recombination dominates, while cooling by C$^+$ and O lines is negligible \cite[][]{2018Haid}.
For heating, we include H$_2$ formation and H$_2$ photodissociation, shock heating and heating due to the release of gravitational potential energy as well as heating by photoionization.
For the latter, each star (see below) is assumed to emit a black body spectrum according to the stellar temperature $T_{\rm int}$. It is used to calculate the mean energy of the ionizing photons, $h\bar{\nu}$, and their average excess energy, $h(\bar{\nu} - \nu_\mathrm{LyC})$, where $\nu_\mathrm{LyC} = 13.6 \, \mathrm{eV}$ is the threshold frequency for hydrogen ionization. 
The photoionization heating rate results in
\begin{eqnarray}
    \Gamma_\mathrm{ih} = n_\mathrm{H}^2 \alpha_\mathrm{B} h (\bar{\nu} - \nu_\mathrm{LyC})
\end{eqnarray}
where $n_\mathrm{H}$ is the hydrogen number density and $\alpha_\mathrm{B}$ is the radiative recombination rate \citep{2018Haid}. 

The equilibrium dust temperature is calculated by taking into account different heating and cooling processes as collisional interactions of dust grains and gas, heating by an interstellar radiation field (here set to 0), heating from H$_2$ formation on grain surfaces, heating by thermal radiation and cooling via thermal emission \citep[see][for details]{2023Kleptiko}.

Sink particles represent protostellar objects and are formed from the gravitational collapse of gas that meets certain criteria \citep[][]{Offner_2009, Klassen_2012,2023Kleptiko}. We use the sink particle implementation of \cite{Dinnbier_2020}, which uses a 4$^{\rm th}$-order Hermite integrator to model their motion. To create a sink particle, the following criteria must be fulfilled. 
Sink particles are formed when the density in a cell is greater than a certain sink density threshold, $\rho_{\rm thresh}$. The sink density threshold is related to the smallest resolvable Jeans length at the maximum refinement level, which is set to $\lambda_{\rm J} = 4 \Delta x$ and can be calculated from
\begin{eqnarray}
    \rho_{\rm thresh} &=& \frac{\pi c_{\rm s}^2}{G \lambda_{\rm J}^2} = \frac{\pi c_{\rm s}^2}{G (4 \Delta x)^2},
\label{eq:rhothresh}
\end{eqnarray}
where $G$ is the gravitational constant and $c_{\rm s}$ is the speed of sound \citep{truelove1997}. 
Within an accretion radius, $r_{\mathrm{acc}} = 2.5 \Delta x$ (= 1000 au for refinement level 9 in the \textsc{Fiducial} run), the gas must be infalling and gravitationally bound. The region lies at a local gravitational potential minimum and does not contain other sink particles \citep{federrath2010}. 
Additionally, a sink can only form from gas that collapses faster than it would accrete onto a nearby sink \citep{2017Clarke}.

Each sink particle is treated as a single star whose radius and surface temperature evolve primarily based on its mass accretion history, following the protostellar evolution model \citep{Offner_2009, Klassen_2012}. The model tracks the star’s properties during the protostellar phase before transitioning to zero-age main sequence (ZAMS), where stars are modelled using the fitting functions of \citep{1996tout}.
The employed stellar evolution model does not depend on the metallicity.
The treatment of the emitted feedback is described in \cite{2023Kleptiko}.  

The emitted luminosity of the sink particles is distinguished into two parts, an internal luminosity, $L_\mathrm{int}$, directly emitted from the star, and its accretion luminosity, $L_\mathrm{acc}$, emitted by accretion events.
For both luminosities, $L_\mathrm{int}$ and $L_\mathrm{acc}$, we calculate the temperatures, $T_\mathrm{int}$ and $T_\mathrm{acc}$, respectively, according to
\begin{eqnarray}
  T_\mathrm{int} &=& \left(\frac{L_\mathrm{int}}{\sigma 4 \pi r^2_\mathrm{star}}\right)^{1/4} \, ,\\
  T_\mathrm{acc} &=& \left(\frac{L_\mathrm{acc}}{f_\mathrm{filling}\sigma 4 \pi r^2_\mathrm{star}}\right)^{1/4} \, ,
\end{eqnarray}
where $r_\mathrm{star}$ is the stellar radius and $f_\mathrm{filling}$ is the fraction of the star's surface that accretes mass.
The accretion luminosity is calculated with a hot spot accretion model and $f_\mathrm{filling}$ is set to 0.1 according to \cite{Calvet1998}. 
Based on the temperatures $T_\mathrm{int}$ and $T_\mathrm{acc}$, we determine the fractions of ionizing radiation ($\gamma(T_\mathrm{int})$ and $\gamma(T_\mathrm{acc})$), and the complementary part of non-ionizing radiation ($[1-\gamma(T_\mathrm{int})]$ and $[1-\gamma(T_\mathrm{acc})]$). Ionizing radiation (UV radiation) are photons emitted in the Lyman continuum, $\nu_\mathrm{LyC}$. Therefore, the fraction of ionizing radiation is given by the luminosity emitted in the Lyman continuum divided by the bolometric luminosity.
\begin{eqnarray}
  \gamma(T_\mathrm{int}) = \frac{\int_{\nu_{LyC}}^\infty \mathbf{d}\nu B_\nu (T_\mathrm{int})}{\int_0^\infty \mathbf{d}\nu B_\nu (T_\mathrm{int})} \, ,
\end{eqnarray}
where $B_\mathrm{\nu}$ is Planck's law of black body radiation. The fraction of ionizing radiation caused by the accretion luminosity is calculated in the same way.
Therefore, for each sink particle, we can distinguish the luminosity of ionizing, $L_\mathrm{UV}$, and non-ionizing, $L_\mathrm{IR}$, radiation:

\begin{eqnarray}
    L_\mathrm{UV} &=& \gamma(T_\mathrm{int})L_\mathrm{int}+\gamma(T_\mathrm{acc})L_\mathrm{acc} \, , \\
    L_\mathrm{IR} &=& L_\mathrm{int} + L_\mathrm{acc} - L_\mathrm{UV} \, .
\end{eqnarray}
 These luminosities are the sources for the radiative transfer described previously.

\begin{table}
\begin{tabular}{l | c | c | c | c | c | c }
     Name & $\rho_0$  & $w$ & $Z$  & $\sigma$ & $\alpha_{\rm vir}$  & $l_{\rm ref}$  \\
      & $[\mathrm{g/cm^3}]$& &$ [\mathrm{Z}_\odot]$ & $ [\mathrm{km/s} ]$ &  &  \\
     \hline\hline
                  \textsc{Fiducial}     & 2.86 $\cdot$ 10$^{-19}$ & -2 &  1&   71.7 & 0.6 & 9 \\
                  \textsc{Rfl10}        &2.86 $\cdot$ 10$^{-19}$ & -2 &  1&  71.7 & 0.6 & \cellcolor{yellow}10 \\
                  \textsc{Rfl11}        &2.86 $\cdot$ 10$^{-19}$ & -2 & 1  & 71.7 & 0.6 & \cellcolor{yellow}11 \\
                  \textsc{Rfl12}        &2.86 $\cdot$ 10$^{-19}$ & -2&  1  & 71.7 & 0.6 & \cellcolor{yellow}12\\
                  \textsc{$\alpha$Low}  &2.86 $\cdot$ 10$^{-19}$ & -2&  1 & \cellcolor{yellow}41.5 &\cellcolor{yellow} 0.2 & 9 \\
                  \textsc{$\alpha$High} &2.86 $\cdot$ 10$^{-19}$ & -2&  1  & \cellcolor{yellow}101.6 &\cellcolor{yellow} 1.2&9\\
                  \textsc{ZLow}         &2.86 $\cdot$ 10$^{-19}$ & -2&  \cellcolor{yellow}0.5  & 71.7 & 0.6&9 \\
                  \textsc{ZHigh}        &2.86 $\cdot$ 10$^{-19}$ & -2& \cellcolor{yellow}2  & 71.7 & 0.6&9 \\
                  \textsc{Flat}         &\cellcolor{yellow} 1.16 $\cdot$ 10$^{-20}$ &\cellcolor{yellow}0& 1  & 71.7 & 0.6&9 \\
                  \textsc{Shallow}      &\cellcolor{yellow}1.60 $\cdot$ 10$^{-19}$ & \cellcolor{yellow}-1.5 & 1  & 71.7 & 0.6&9  \\

\end{tabular}
\caption{Initial conditions of the performed simulations. From left to right, the columns show the run name, central density $\rho_0$, powerlaw exponent of the initial core density profile $w$, core radius $r$, metallicity $Z$, velocity dispersion $\sigma$, virial parameter, $\alpha_{\rm vir}$ and the maximum refinement level $l_{\rm ref}$. With respect to the \textsc{Fiducial} run, the yellow cells indicate the parameter changed in each row. }
\label{tab_parameter}
\end{table}

\subsection{Initial Conditions}
\label{Sec. Initial Conditions}

The initial core setups are designed to ensure the formation of massive stars as the cores collapse under self-gravity. These stars grow by accreting mass until their own feedback halts further accretion, ultimately determining their final mass.

The initial conditions for our study, including density profile, virial parameter and metallicity are motivated by observational data from the ALMAGAL survey of clumps and cores (\citep{2025Molinari, 2025Sanchez, 2025Coletta}. This survey provides a representative sample of star-forming regions with measured physical properties, allowing us to select parameter values that reflect typical environmental conditions. By grounding our parameter choices in this observational dataset, we ensure that our simulations explore realistic regimes relevant to massive star formation.

All cores have a mass of $1000 \, \mathrm{M}_\odot$ within a radius of $1 \, \mathrm{pc}$, which could be considered a typical setup for a massive star-forming core \citep{2014Tan}. The initial gas temperature is $20 \, \mathrm{K}$ and the initial dust temperature of $2.7 \, \mathrm{K}$ is immediately adjusted to the thermal equilibrium value in the first time step.
The side length of the cubic box is 4~pc and the cores are surrounded by a low density ambient medium.
The parameters we vary are the density profile, the virial parameter, and the metallicity (see Table \ref{tab_parameter}).

\subsubsection{Density Distribution}
For the initial density distribution we use a Plummer-like density profile, i.e.
\begin{eqnarray}
    \rho(r) &=& \frac{\rho_0}{1+(\frac{r}{r_0})^w}\, ,
\end{eqnarray}
where $r$ is the core radius, $\rho_0$ the central density, $w$ the density power-law, and $r_0$ the scale radius ($\sim 0.15 \rm pc$). Three different density power-laws are used, $w=2$, $w=1.5$, and $w=0$, where $w=0$ results in a constant density profile with $\rho = \rho_0$. Assuming a constant density distribution the core has a Jeans length of $r_{_\mathrm{J}} = 0.26\, \mathrm{pc}$, resulting in a Jeans mass of $m_{_\mathrm{J}} = 18.1 \, \mathrm{M}_\odot$. 
The core is pressure-confined by the environment; therefore, at the core boundary the density decreases by a factor of $100$ while the temperature of the surrounding medium is increased to $2000 \, \mathrm{K}$.

Assuming a constant density $\bar{\rho}$, the free-fall time, $t_{\mathrm{ff}}$, is the characteristic time it would take for a core to collapse purely under its own gravity. It is given by:
\begin{eqnarray}
\label{eq:tff}
    t_{\mathrm{ff}} =  \sqrt{\frac{3 \pi}{32 G \bar{\rho}}}\, .
\end{eqnarray}
In our simulations the free-fall time is $t_{\mathrm{ff}} = 0.526$~Myr. We normalize multiple time evolution plots to this $t_{\mathrm{ff}}$.

\subsubsection{Turbulence and Virial Parameter}
For the initial turbulent velocity field, we generate a random Gaussian velocity field \citep{BurkertBodenheimer2000}. It follows a power-law spectrum $P_\mathrm{k} \propto k^{-4}$ (Burgers turbulence in 3D), while using the minimum and maximum wave numbers $k_\mathrm{min} = 1$ and $k_\mathrm{max} = 64$ in Fourier space, where $k=1$ corresponds to $2 \pi/ 4pc$. The velocity field is scaled to the desired root-mean-square velocity and mapped onto a uniform $128^3$ grid.

To investigate the impact of turbulence on the cloud evolution we change the virial parameter, $\alpha_\mathrm{vir}$. The virial parameter is determined by the ratio of kinetic and gravitational energy.
The velocity field is normalized; thus, we can derive different turbulent field strengths by multiplying with a chosen velocity dispersion. 
The velocity dispersion, $\sigma$, changes with the virial parameter according to
\begin{eqnarray}
    \sigma = \sqrt{\frac{\alpha_{\mathrm{vir}} G M}{5r}},
\end{eqnarray}
where $M$ is the total core mass. We change the virial parameter from subvirial with a very low velocity dispersion ($\alpha_\mathrm{vir} =0.2$) to supervirial ($\alpha_\mathrm{vir} =1.2$).

\subsubsection{Metallicity}
The fiducial setup has solar metallicity ($Z = 1 \, \mathrm{Z}_\odot$) with the according dust mass fraction, $f_\mathrm{d}$ (see Eq. \ref{eq:fd}). The abundances of metals in the gas-phase as well as $f_\mathrm{d}$ are reduced or increased linearly with $Z$ to model different metallicities. This affects the ability of the gas to cool and shield. We vary the metallicity to $0.5 \, \mathrm{Z}_\odot$ and $2 \, \mathrm{Z}_\odot$ to mimic the conditions of massive star-forming cores towards the outer Milky Way disc and near the Galactic Centre.

\section{Analysis}
\label{sec:Analysis}

\begin{figure*}
	\includegraphics[width=\textwidth]{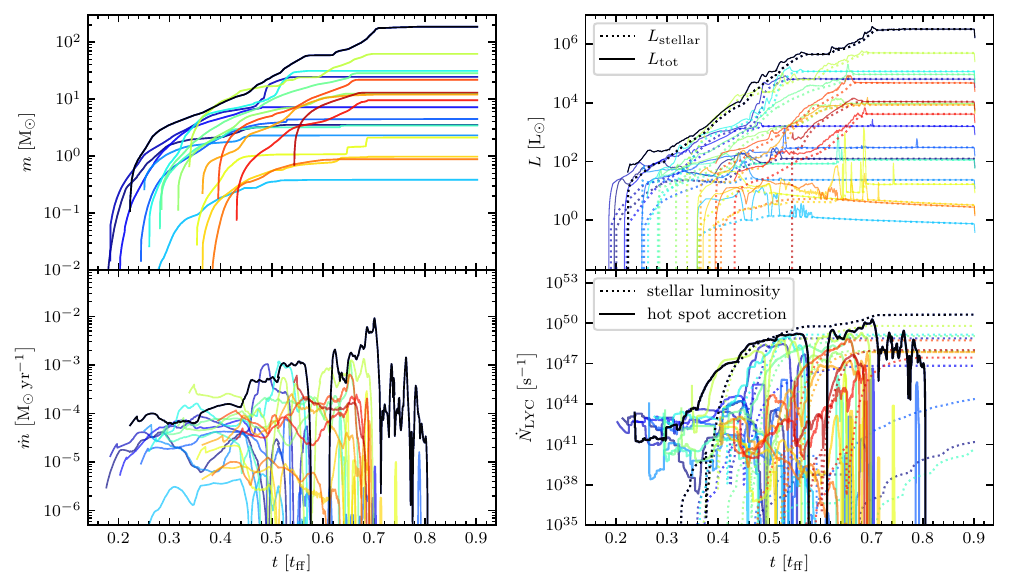}
    \caption{Properties of sink particles in time in the \textsc{fiducial} run. 
    \texttt{Top Left}: Sink mass.
     \texttt{Top Right}: Luminosity (smoothed). 
     \texttt{Bottom Left}: Accretion rate (smoothed).
     \texttt{Bottom Right}: Rate of emitted ionizing (Lyman Continuum) photons (smoothed).
    In total 20 sink particles are formed. The sink particle that ends up being most massive (MMS) is coloured in black.}
    \label{fig:sinkprop}
\end{figure*}

\begin{figure*}
	\includegraphics[width=\textwidth,trim=0 0.2cm 0.0cm 0cm, clip]{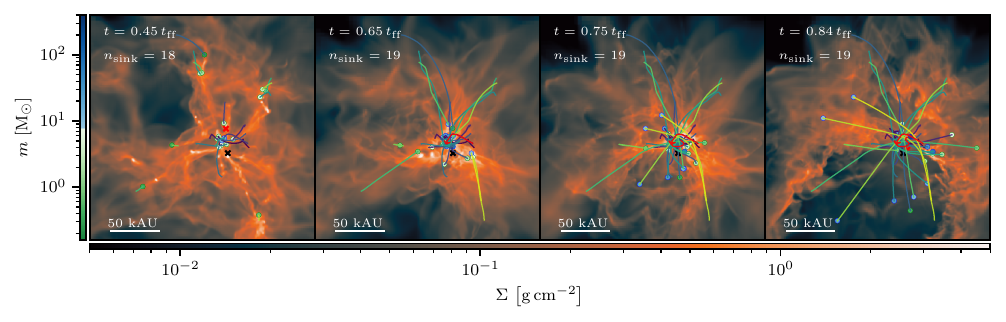}
    \caption{Time evolution of the column density of the \textsc{Fiducial} run. The circles show the current position of the sink particles, similar to Fig. \ref{fig:fiducialrun}. The lines show the track of each sink particle, coloured as in Fig. \ref{fig:sinkprop}. The geometric centre and the centre of mass are marked with black and red crosses, respectively. Sink particles that are formed in the outer region fall towards the central region and immediately escape the potential well due high velocities.}
    \label{fig:fiducialCDsinks}
\end{figure*}

\begin{figure}
	\includegraphics[width=\columnwidth]{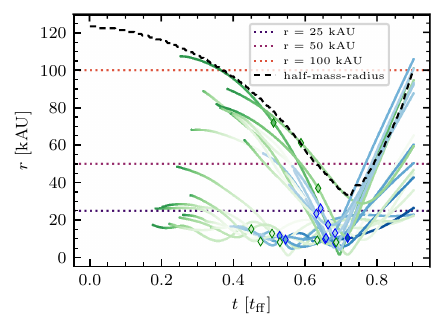}
    \caption{Evolution of the radial distribution of sink particles with respect to the geometric centre of the simulation box. The colour code shows the mass of the sink particles, while diamonds indicate when the sink particle has accreted $95\%$ of its final mass. The black dashed line shows the half-mass-radius, which is the radius of the sphere in which half of the initial mass (here $500 \, \mathrm{M}_\odot$) is located. The dotted lines indicate the radii used in Fig. \ref{fig:massoutflow}.
    Sink particles that are formed near the geometric centre stay near the centre, while particles that are formed outside perform a V-shaped movement in terms of radius.
    The MMS is formed near the centre, while sinks with very low final masses are formed at larger distances. However, for intermediate final masses, sink particles can form near the centre or in the outer parts. }
    \label{fig:radialDistribution}
\end{figure}

\begin{figure}
	\includegraphics[width=\columnwidth]{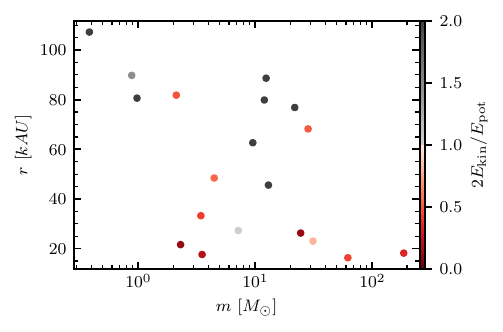}
    \caption{Radial distance of the sink particles at the formation time with respect to the geometric centre compared to their final mass. The colour shows if a sink particle is gravitationally bound (red) or unbound (grey) after $0.9 t_\mathrm{ff}$ when the parental cloud is dispersed. 10 out of 19 sink particles remain gravitationally bound at this point. We find that it is more likely that a sink particle ends up staying bound if it has been formed close to the centre.}
    \label{fig:radius_mass_bound}
\end{figure}

\begin{figure}
	\includegraphics[width=\columnwidth]{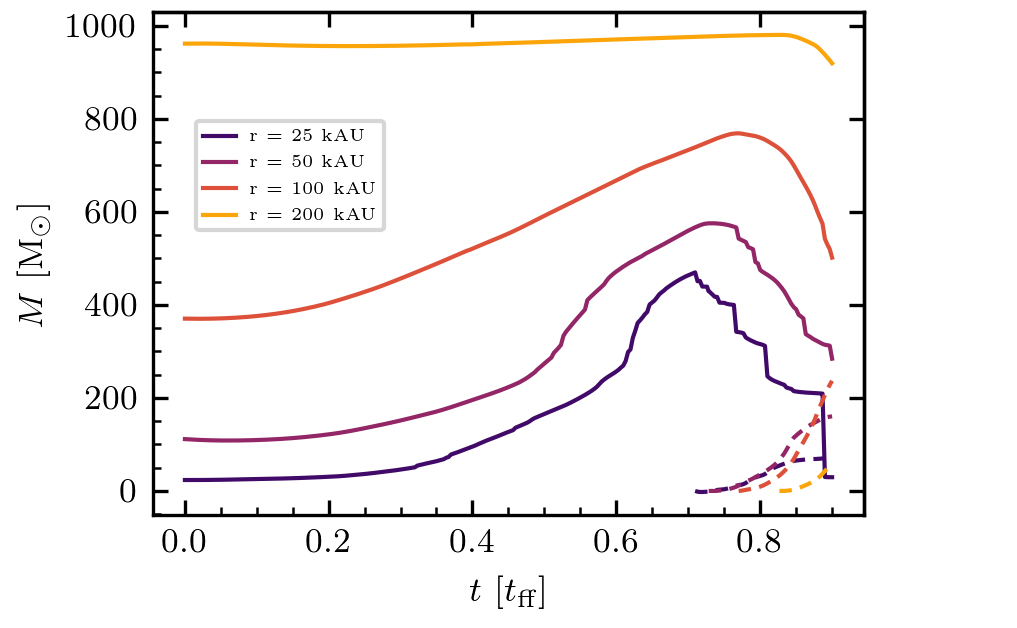}
    \caption{The solid line shows the mass evolution within spheres with different radii around the geometric centre (masses of gas and sink particles are considered). The dashed lines show the outflowing gas mass that is leaving these spheres due to feedback. First, the mass of the inner region of the cloud increases due to the gravitational collapse and later decreases again when either sink particles leave the region (sharp decreases) and/or feedback blows material out. }
    \label{fig:massoutflow}
\end{figure}

\begin{figure*}
	\includegraphics[width=0.8\textwidth,trim=0 0 0.5cm 0.0cm]{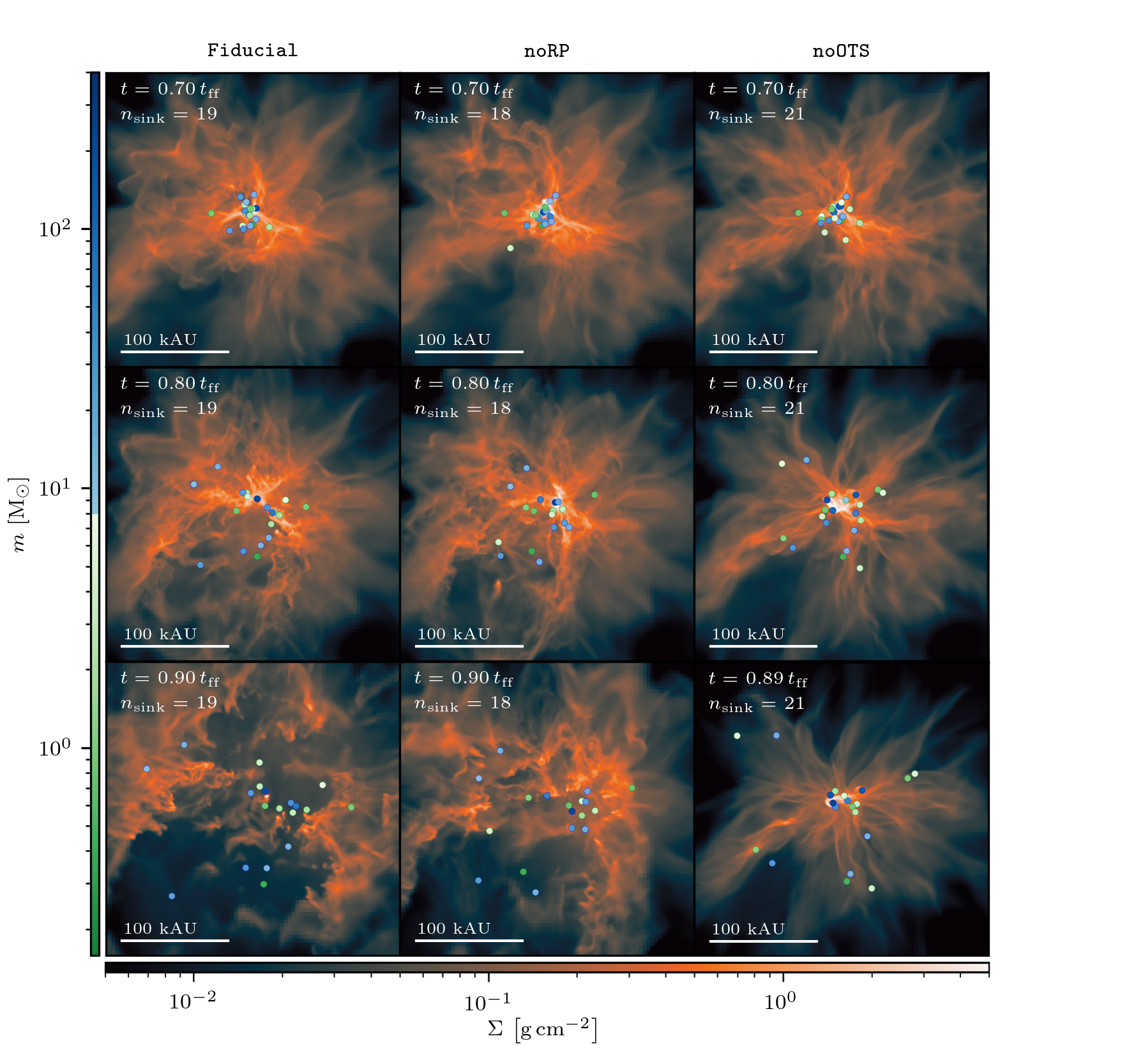}
    \caption{Time evolution (from top to bottom) of the column density of simulations with different feedback mechanisms. From left to right we plot the \textsc{Fiducial} run (both ionizing radiation and radiation pressure are included), the simulation without radiation pressure from ionizing radiation (\textsc{noRP}) and the run without ionizing radiation at all (\textsc{noOTS}).
    Ionizing radiation becomes important as soon as sink particles reach $8 \,\mathrm{M_{\odot}}$ and more. Radiation pressure of ionizing radiation transfers momentum to gas and dust and supports the grow of the HII region.
    }
    \label{fig:combinedRadiation}
\end{figure*}

\subsection{Fiducial Run}

We investigate the formation and early evolution of massive stars and their host cores up to the point where UC-HII regions are established (see Fig. \ref{fig:volren}). Here, we discuss the \textsc{Fiducial} run. 
\subsubsection{General Evolution}
We first present the time evolution (top to bottom) of the gas column density, $\Sigma$, and the mass-weighted temperatures of gas, dust, and radiation, $T_\mathrm{gas}$, $T_\mathrm{dust}$, and $T_\mathrm{rad}$, respectively (see Fig. \ref{fig:fiducialrun}).
The cloud core is initialized to collapse under its own gravity. Due to the interplay of gravitational and turbulent kinetic energy, substructures form and evolve into several filaments. The filaments converge at the central hub. The first sink particles are formed in this central hub, whereas later on sink particle formation also takes place in the outer parts of the filamentary structures. 

At first, the temperatures of dust and gas show a similar morphology as $\Sigma$ (see Fig. \ref{fig:fiducialrun}). The first sink particle is formed after $0.18 \,\mathrm{t_{ff}}$ (see Eq.~\ref{eq:tff} and Table~\ref{tab_parameter}). As a result of the formation of sink particles and their growing feedback, the dust is heated by radiation. After $\sim 0.4 \, \mathrm{t_{ff}}$ the first massive sink particles develop. From this point on, the dust temperature is dominated by the radiation temperature. 
In the inner region, the temperatures of gas, dust, and radiation are mostly in equilibrium. However, in the outer region, the temperature of the gas is still higher due to shock heating. Later, ionizing radiation from massive sink particles heats the gas and the UC-HII region forms and expands. As a consequence, the gas temperature increases significantly.
A phase diagram of the temperature evolution of gas, dust, and radiation temperature according to Fig.\ref{fig:fiducialrun} is shown in Fig. \ref{fig:fiducialrunTemp}. 

\subsubsection{Sink Particle Evolution}
Figure \ref{fig:sinkprop} illustrates the properties of the sink particle population. The time evolution of the mass (top left panel), accretion rate (bottom left), luminosity (top right) and ionizing photon rate (bottom right) are plotted. The most massive sink particle (MMS), defined as the sink that is most massive at the end of the simulation, is shown in black.
For better readability, the accretion rate, ionizing photon rate, and luminosity derived from sink particles in this work are smoothed using a moving average with a window size of 250 data points.
The formation of sink particles takes place from $0.18 \,\mathrm{t_{ff}}$ to $0.55 \, \mathrm{t_{ff}}$. 
The first formed sink particle does not evolve into the most massive one. The MMS is built from the fourth particle, which forms after $0.22 \, \mathrm{t_{ff}}$ and is located close to the centre of the core.
The MMS reaches a final mass of $187.2 \, \mathrm{M_{\odot}}$ at $0.8 \, \mathrm{t_{ff}}$ (Fig.~\ref{fig:sinkprop}, top left panel). 
We note that the MMS does not necessarily correspond to a single star. Although we treat it as such in this work, it may host multiple stars, as we only take into account turbulent fragmentation and no further disc fragmentation.
(see section \ref{Sec:ResolutionStudy} for more details). A binary (or multiple system) would most likely provide less feedback than a single star with the same mass. 
Additionally, \cite{2012Hosokawa} report that a protostar's radius may oscillate, and its averaged effective temperature should decrease with an accretion rate higher than $3 \times 10^{-3}  \, \mathrm{M_{\odot}/yr}$. Therefore, we might somewhat overestimate the properties and (ionizing) feedback from the massive sink particles. 

Fragmentation plays a crucial role in our simulation. Turbulence drives fragmentation, leading to the formation of multiple stars, each contributing feedback against the infalling gas from the surrounding core. This feedback compresses the gas, halting accretion for some stars and enhancing it for others. As a result, the turbulent accretion process produces highly variable accretion rates. The interplay between local and global collapse ultimately leads to the formation of the MMS.

The average accretion rate of the MMS is $\sim 1 \times 10^{-3}  \, \mathrm{M_{\odot}/yr}$ and peaks at time $0.69 \, \mathrm{t_{ff}}$ with $2.9 \times 10^{-2}  \, \mathrm{M_{\odot}/yr}$, just before the feedback begins to suppress further accretion (Fig. \ref{fig:sinkprop}, bottom left panel). 
Between $0.56 \, \mathrm{t_{ff}}$ and $0.61 \, \mathrm{t_{ff}}$, accretion onto the MMS is partially prevented. Ionizing feedback from the sink itself pushes the gas away and locally stops accretion. However, the internal feedback from the MMS is not strong enough to completely stop further accretion. Because the global collapse continues, after $0.05 \, \mathrm{t_{ff}}$, the gravitational energy is again stronger than the feedback and the local collapse continues.

In the top right panel of Fig. \ref{fig:sinkprop}, the stellar and total luminosity is shown, while the total luminosity is just the sum of the stellar and the accretion luminosity. 
The stellar luminosity is the main contribution to the total luminosity. It increases continuously, and the MMS reaches $3.2 \times 10^6 \, \mathrm{L_{\odot}}$. The accretion luminosity gives an additional noisy contribution to the total luminosity if accretion events are happening. 

Some ionizing photons (bottom right panel) are already emitted before the first massive star is born because of the hot spot accretion. 
The ionizing photon rate, $\dot{N}_{\mathrm{LyC}}$, gradually increases before sharply rising once the first sink particle reaches $8 \, \mathrm{M}_\odot$.
The MMS reaches a maximum value of $\dot{N}_{\mathrm{LyC}} = 4 \times 10^{50}\, \mathrm{s}^{-1}$. For comparison, a typical O-type star with around $25 \, \mathrm{M}_\odot$ would have an average value of $\dot{N}_{\mathrm{LyC}} \sim 10^{49}\, \mathrm{s}^{-1}$ \citep{2005Tielens}.
During the aforementioned accretion pause of the MMS only ionizing photons from the star itself are emitted while accretion does not contribute any photons at this time.

\subsubsection{Sink Particle Dynamics}
Fig. \ref{fig:fiducialCDsinks} shows the movements of the sink particles in time on top of the column density. The colours of the tracks are the same as in Fig. \ref{fig:sinkprop}. The sink particles that are formed in the outer region initially fall towards the centre. Because of their high radial velocities and a decreasing potential well depth, they almost immediately escape the potential well again.
When a substantial amount of ionizing photons is first emitted (the first stars become massive after $\sim 0.4 t_{\rm ff}$), the dense gas near the central sink-forming hub first hinders the formation of an UC-HII region. 
In total 10 sink particles reach $> 8 \, \mathrm{M}_{\odot}$ (see section \ref{sec:ParameterStudy}).
As soon as massive sink particles either move into or reside in lower-density regions because they have eaten most of their surrounding gas, the radiative feedback is more and more able to push the gas away from the centre.
Thus, the UC-HII region around the massive sink particles expands and the cavity grows ($\sim 0.7 t_{\rm ff}$).
For some sink particles, accretion stops completely at this point, while the MMS sink particle is still sitting in a dense hub where its radiation is trapped and mass accretion is still ongoing. Nevertheless, the radiation from nearby massive stars (although less massive than the MSS), which live in somewhat more diffuse environments, keeps pushing gas out of their surroundings and starts to disrupt the inner hub. The MMS continues to accrete until $\sim 0.8 t_{\rm ff}$.
In the end, the remaining gas is dispersed. 

The sink particles that accumulated high radial velocities at formation are now moving outwards.
A more detailed analysis of the star formation rate and feedback is done in the following.

Fig. \ref{fig:radialDistribution} shows the time evolution of the radial distribution of stars. In particular, we plot the radial distances between the positions of each sink and the geometric centre of the simulation box. Particles that are formed near the central region stay near the centre until the remaining cloud is dispersed. 
Sink particles that are formed outside perform a V-shaped movement in terms of radius in time. They enter the central region and immediately leave again. The MMS is formed at $20 ~ \mathrm{kau}$ from the centre. Generally, we find that sink particles with low final masses most likely form at larger distances, while sink particles with intermediate final masses form either near the centre or in the outer parts. The diamond-shaped symbols indicate the time that a sink has accreted 95\% of its final mass. It can be seen that the sink particle accretion is basically complete before the cluster begins to disperse at $\sim 0.7 \, \mathrm{t_{ff}}$.

The black dashed line represents the radius of a sphere around the geometric centre, which contains half the mass of the initial cloud, i.e., the sphere within the half-mass-radius (HMR) contains $500 \, \mathrm{M}_\odot$. The HMR is initially $123\, \mathrm{kau}$. During collapse, it decreases to $32 \, \mathrm{kau}$ and increases again as soon as the cluster is dispersed. All sink particles are formed within the HMR.

The evolution of the sink particle cluster is determined by N-body dynamics and the gravitational interaction of the sinks and the gas distribution. Overall, we show that not all sink particles remain in a gravitationally bound cluster. Fig. \ref{fig:radius_mass_bound} depicts whether or not a sink particle is bound after $0.9 \, t_\mathrm{ff}$. We find that 10 out of 19 sink particles end up in a gravitationally bound system. With a total SFE of $42 \, \%$ (discussed in section \ref{sec:ParameterStudy}) around $53 \, \%$ of the sink particles remain bound in a cluster, which agrees with earlier studies \citep{2000Adams,2007Baumgardt}. If a sink particle is either very massive (here $>25 \, M_\mathrm{\odot}$) or formed close to the centre it is more likely that it stays in a bound system.

\subsubsection{Gas Dispersal and Outflowing Mass}
Stellar feedback drives outflows, resulting in the dispersal of the core.
Fig. \ref{fig:massoutflow} shows the evolution of the total mass within spheres with different radii around the geometric centre. First, the mass at the centre increases due to the gravitational collapse. In the inner region ($r = 25 \, \rm{kau}$) almost all gas is accreted onto sink particles.
After $\sim 0.7 \, \mathrm{t_{ff}}$, feedback becomes dominant and starts to push the gas out. Therefore, the mass begins to decrease near the centre. This also coincides with the time when the sink particles begin to move outwards (see Fig.~\ref{fig:radialDistribution}, dotted lines indicates the radii used in Fig. \ref{fig:massoutflow}).
The dashed lines in Fig. \ref{fig:massoutflow} represent the gas mass, which leaves the boundary of the respective sphere. At a radius of $100 \, \rm{kau}$ over $230 \, \mathrm{M}_\odot$ have been blown out of the sphere after $0.9 \, t_\mathrm{ff}$. The decrease in mass occurs with a time delay at larger radii, indicating an inside-out expansion.

\begin{table}
	\centering
	\begin{tabular}{lcccc}
		\hline
		Name &  ionizing &  non-ionizing & ionizing    \\
		     &  radiation &  radiation    & rad. pres. \\
		\hline
		\textsc{Fiducial}  & x & x & x  \\
		\textsc{noRP}      & x & x &  \\
		\textsc{noOTS}    &   & x &    \\
		\hline
	\end{tabular}
	\caption{\label{tab:radiation} List of runs investigating the impact of ionizing radiation and radiation pressure}
\end{table}

\begin{figure}
	\includegraphics[width=\columnwidth]{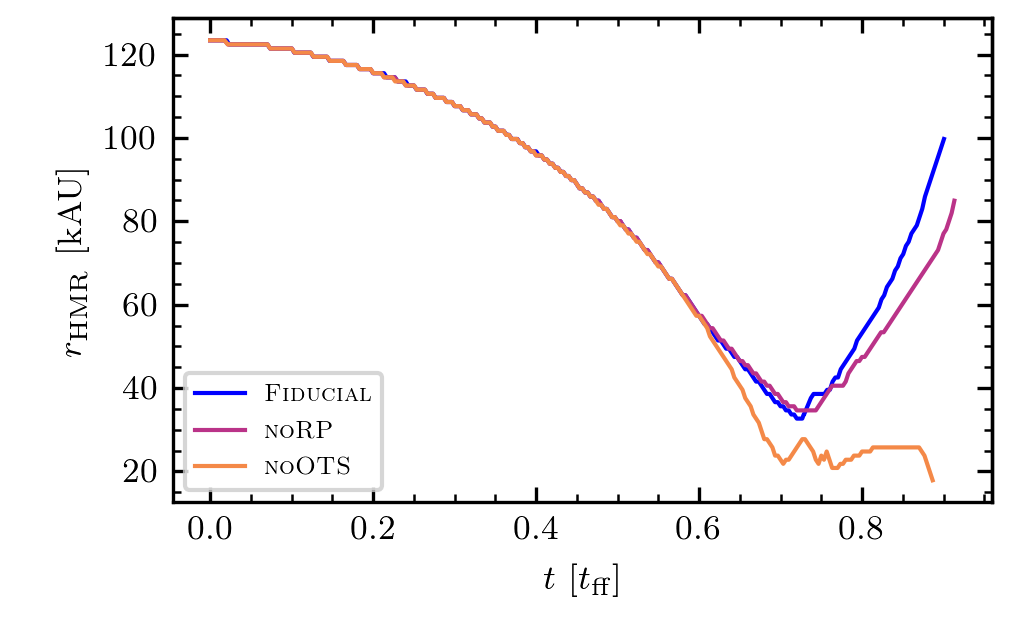}
    \caption{Evolution of the half-
mass-radius, $r_\mathrm{HMR}$, which is the radius of the sphere in which half of the initial mass
($500 \, M_\odot$) is located, showing the expansion of the cloud core. Ionizing radiation ultimately stops the core collapse and causes the cloud to re-expand. The extra momentum from RP accelerates the expansion.\\}
    \label{fig:radiationHMR}
\end{figure}

\begin{figure*}
     \centering
     \begin{subfigure}[b]{0.45\textwidth}
         \centering
         \includegraphics[width=\columnwidth,trim=0 0 0 0.056cm, clip]{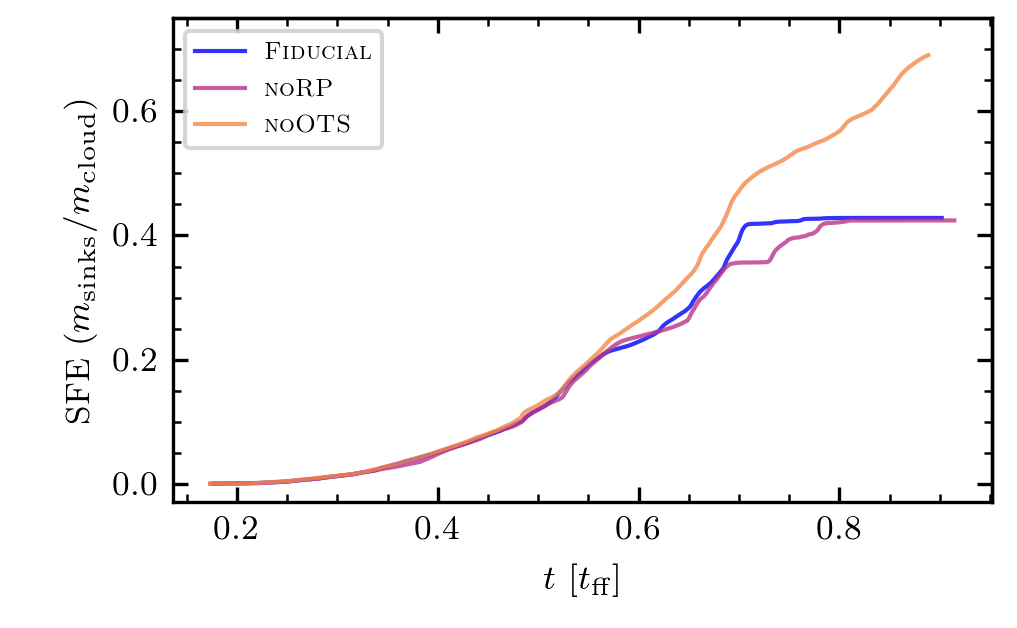}
     \end{subfigure}
     \hfil
     \begin{subfigure}[b]{0.45\textwidth}
         \centering
         \includegraphics[width=\columnwidth,trim=0 0 0 0.056cm, clip]{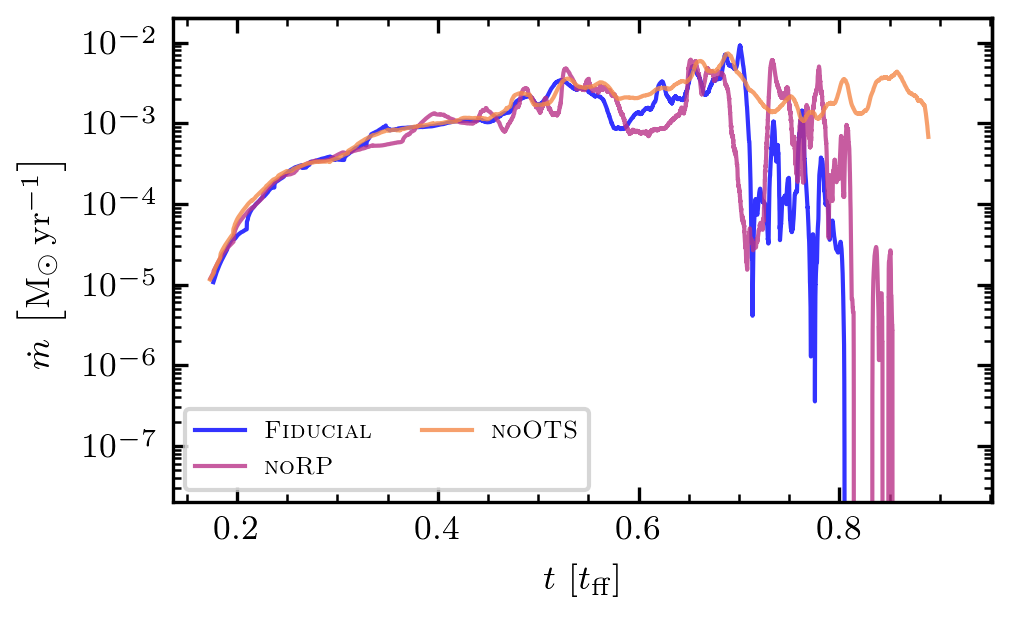}
     \end{subfigure}\\
     \begin{subfigure}[b]{0.45\textwidth}
         \centering
         \vspace{-1.04cm}
         \includegraphics[width=\columnwidth,trim=0 0 0 0.15cm, clip]{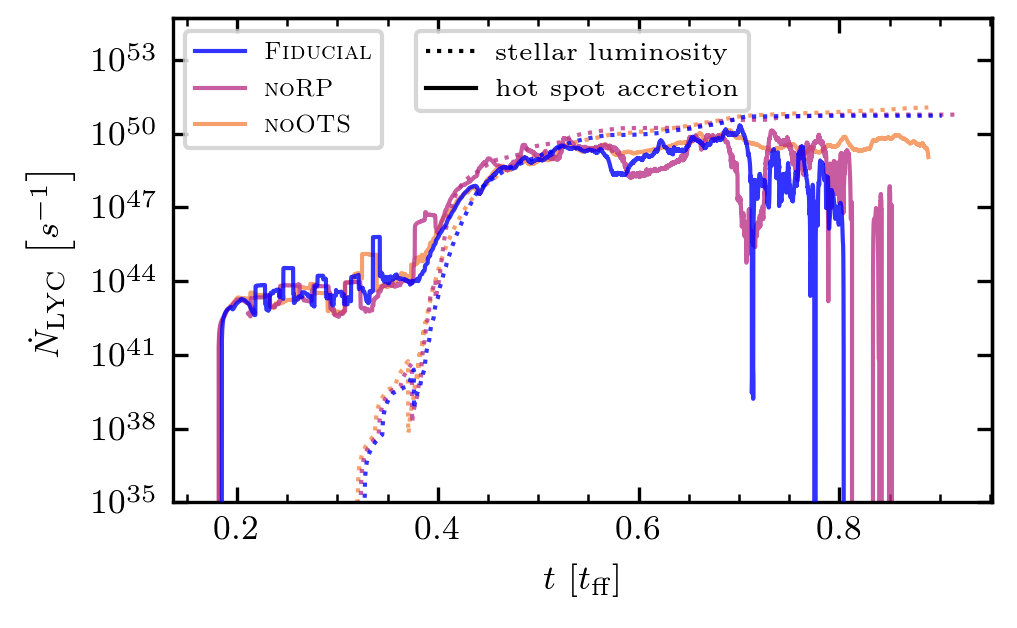}
     \end{subfigure}
     \hfil
     \begin{subfigure}[b]{0.45\textwidth}
         \centering
         \vspace{-1.04cm}
         \includegraphics[width=\columnwidth,trim=0 0 0 0.15cm, clip]{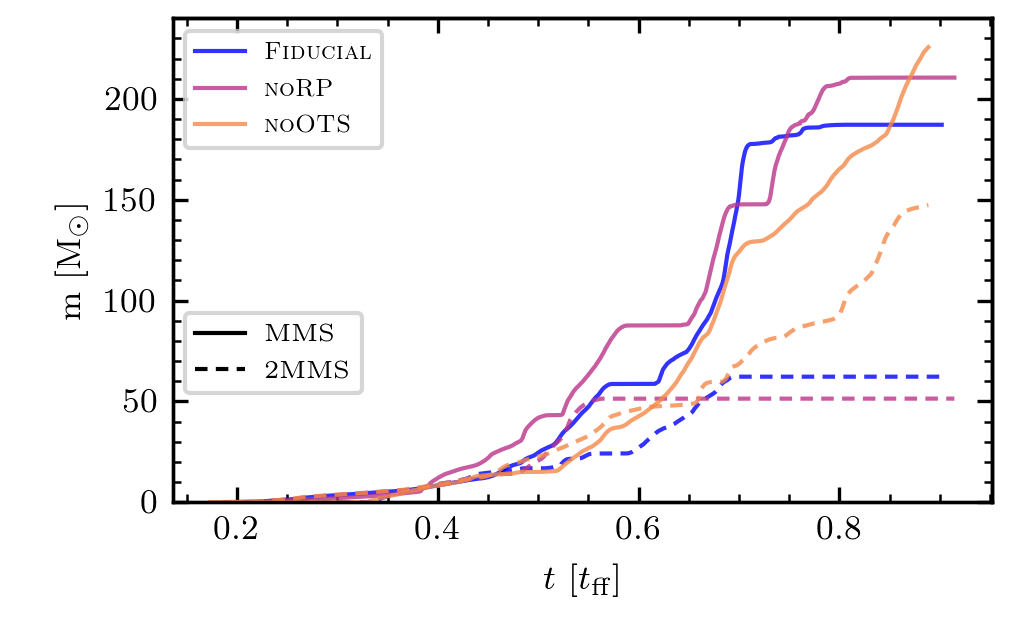}
     \end{subfigure}
     \caption{\texttt{Top Left}: Evolution of the star formation efficiency.
     \texttt{Top Right}: Total accretion rate (smoothed). 
     \texttt{Bottom Left}: Total ionizing photon rate (smoothed).
     \texttt{Bottom Right}: Mass of the most massive sink (MMS) and the second most massive sink (2MMS).
     Ionizing radiation suppresses mass accretion and finally prevents the mass growth of sink particles completely. For runs without ionizing feedback, mass accretion onto sink particles is not prevented and the SFE increases until the simulation is stopped. 
     }
     \label{fig:radiationSFE}
\end{figure*}

\begin{figure*}
     \centering
     \begin{subfigure}[b]{0.45\textwidth}
         \centering
         \includegraphics[width=\columnwidth,trim=0 0 0 0.056cm, clip]{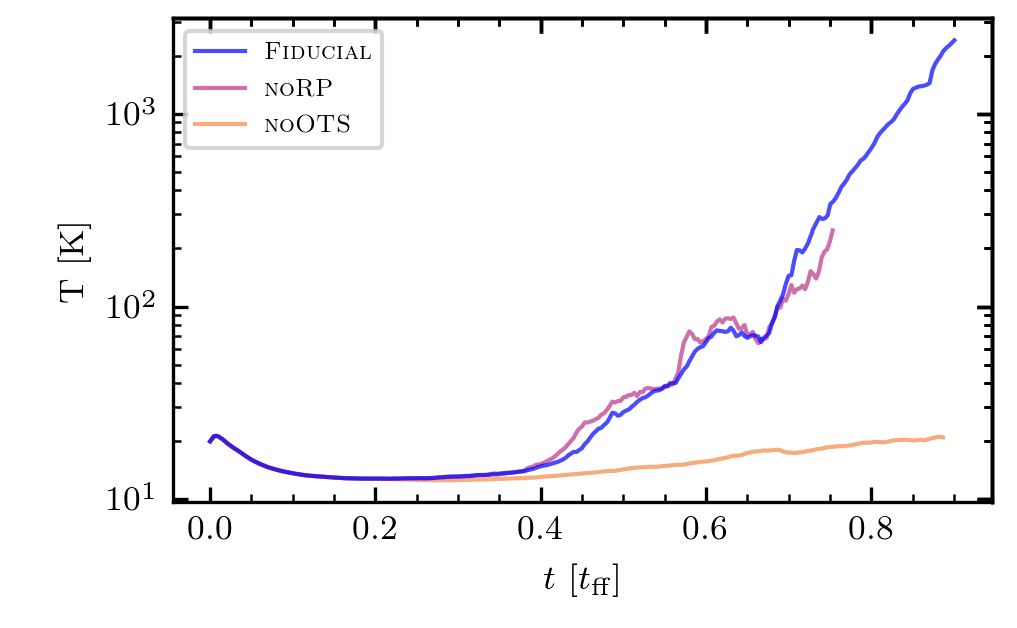}
     \end{subfigure}
     \hfil
     \begin{subfigure}[b]{0.45\textwidth}
         \centering
         \includegraphics[width=\columnwidth,trim=0 0 0 0.056cm, clip]{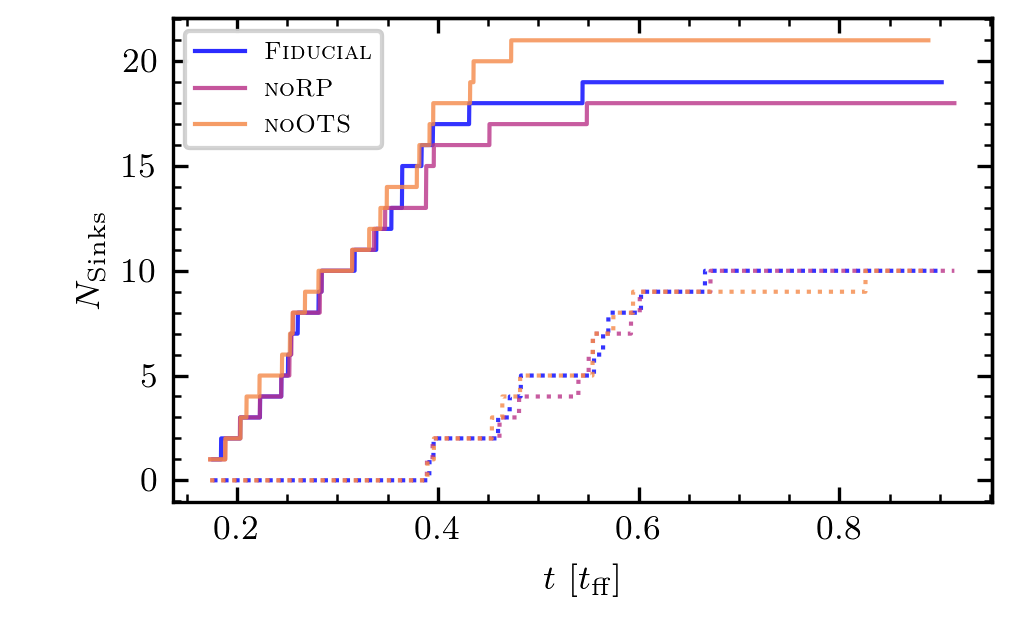}
     \end{subfigure}\\
     \caption{\texttt{Left}: Mass-weighted gas temperature of the whole simulation domain.
     \texttt{Right}:Number of sink particles, while the dotted line shows the number of high mass sinks with $\geq 8M_\odot$.    
     Due to ionizing radiation, the temperature starts to heat up after  $\sim 0.4 t_{\mathrm{ff}}$, while simulations without ionizing radiation have a constant shallow temperature increase. The increase in temperature results in a decrease in the number of lower-mass sink particles.}
     \label{fig:MwgTradiation}
\end{figure*}

\subsection{Impact of Ionizing Radiation and Radiation Pressure}
\label{RPandIR}
In this section, we discuss the impact of ionizing radiation and radiation pressure from ionizing radiation on our core collapse scenario. 

We compute three simulations with the same initial conditions but evolving them with different radiation physics (see Tab. \ref{tab:radiation}). The first run is the \textsc{Fiducial} run, which includes ionizing radiation and RP. 
Secondly, we run the simulation without RP (\textsc{noRP}). Then we neglect both, ionizing radiation and RP (\textsc{noOTS}). Heating by non-ionizing radiation from sinks and dust is always included (see sec. \ref{sec:NumMeth}).

\subsubsection{Morphological Evolution}
In Fig. \ref{fig:combinedRadiation} the time evolution of the column densities is shown for the three simulations with different feedback mechanisms. 
The initial cloud collapse phase is similar for all simulations and noticeable differences only emerge once the sink particles become massive enough for feedback to become efficient (after $\sim 0.4 t_{\rm ff}$; see Fig.~\ref{fig:sinkprop}). For this reason, the early evolution is not shown from Fig. \ref{fig:combinedRadiation}.
The first two columns in Fig. \ref{fig:combinedRadiation} (\textsc{Fiducial} run and \textsc{noRP}) include ionizing radiation.
Ionizing feedback becomes significant once sink particles exceed $8 \,\mathrm{M_{\odot}}$, ionizing the surrounding hydrogen within the Str\"omgren radius and forming an UC-HII region.
If no ionizing radiation is included (Fig.~\ref{fig:combinedRadiation}, last column, run \textsc{noOTS}), no gas is pushed outwards and the cloud still collapses when the simulation is stopped (also see Fig. \ref{fig:radiationHMR}. 

\subsubsection{Role of Radiation Pressure in UC-HII Expansion}

When RP is included, the absorption of ionizing photons
is associated with a momentum transfer. In particular, near the sink particles, the radiation energy density is high. As a result, the momentum input by RP pushes the gas away, which reduces the local density. Consequently, the recombination rate within the ionized gas is also reduced, since it depends on the square of the local gas density. Thus, the UC-HII region in the \textsc{Fiducial} run can expand faster compared to \textsc{noRP} (see Fig. \ref{fig:combinedRadiation}, bottom panels, and Fig. \ref{fig:radiationHMR}). The total number of ionizing photons, both, from stellar luminosity and from accretion events, is comparable for both runs, \textsc{Fiducial} and \textsc{noRP} (see Fig. \ref{fig:radiationSFE}, bottom left), which indicates that the UC-HII region expansion may be promoted by RP.
We note here that a different spatial distribution of (massive) sink particles may also contribute to a different shape of the UC-HII region in \textsc{noRP}.

\subsubsection{Star Formation Efficiency and Accretion Rates}

Previous studies show that ionization feedback is effective on suppressing star formation in giant molecular clouds \citep[e.g.][]{2016Geen, 2018Kim, 2020Fukushima}. In runs \textsc{Fiducial} and \textsc{noRP}, the feedback and pressure of ionizing radiation blow out the gas, thereby suppressing further mass accretion onto sink particles. This can be seen in Fig. \ref{fig:radiationSFE}, top left panel, where the slope of the SFE becomes flatter after $0.55 \,\mathrm{t_{ff}}$ when the UC-HII region starts to grow. Feedback from massive sink particles is strong enough to finally fully prevent further accretion (Fig. \ref{fig:radiationSFE}, top right panel, showing the total mass accretion rate onto sinks), so that the SFE stays constant after around $0.8 \,\mathrm{t_{ff}}$ for both runs, \textsc{Fiducial} and \textsc{noRP} (Fig.~\ref{fig:radiationSFE}, top left panel).

The mass accretion of the run \textsc{noOTS} is comparable to the \textsc{Fiducial} run but does not decrease after $\sim 0.7 \,\mathrm{t_{ff}}$ and stays roughly constant at a few times $10^{-3}\,{\rm M}_{\odot}\,{\rm yr}^{-1}$ even at late times (Fig. \ref{fig:radiationSFE}, top right panel). After the initial growth phase, the SFE of the run \textsc{noOTS} increases linearly and reaches $\sim 0.65$ during the simulation time, while there is no indication to stop further growth for longer simulation times (Fig. \ref{fig:radiationSFE}, top left panel). This indicates that ionizing radiation is the most efficient feedback mechanism in our simulations, assisted by RP.

The mass accretion of the MMS of run \textsc{noOTS} is slightly time-delayed but, in the end, the MMS grows even more massive than in the \textsc{Fiducial} run, as there is no mechanism to stop further mass accretion (see Fig. \ref{fig:radiationSFE}, bottom right panel).

\subsubsection{Thermal Impact and Fragmentation}

Ionizing feedback (runs \textsc{Fiducial} and \textsc{noRP}) leads to an increase of the mass-weighted temperature after $\sim 0.4 \,\mathrm{t_{ff}}$ as soon as the first sink particle becomes massive and starts to emit ionizing radiation (see Fig. \ref{fig:MwgTradiation}, left panel). In comparison, run \textsc{noOTS} shows a significantly cooler core with a gradually increasing average temperature. The difference in the temperature evolution affects the fragmentation process. 
The fragmentation process follows a similar pattern in all runs up to $\sim 0.4 \,\mathrm{t_{ff}}$.
As the temperature increases, the Jeans mass increases, and fragmentation becomes more demanding. This leads to a slightly lower final number of sink particles in the \textsc{Fiducial} run and \textsc{noRP} (see Fig. \ref{fig:MwgTradiation}, right panel).
In run \textsc{noOTS} the formation process is quicker (until $\sim 0.47 \,\mathrm{t_{ff}}$) but leads to the formation of more sink particles.
Nevertheless, the number of high-mass sinks (dotted lines on the right panel) stays almost the same for all runs.

\begin{table}
    \centering
   \begin{tabular}{l c c c }
			Name & $l_{\rm ref}$ & $\Delta x$ [au] & $\rho_{\rm thresh}$ [g/cm$^3$]\\
			\hline \hline
			\textsc{Fiducial} & 9 & 400 & $3.67 \times 10^{-17}$ \\  
			\textsc{Rfl10} & 10 & 200 & $1.48 \times 10^{-16}$ \\ 
			\textsc{Rfl11} & 11 & 100 & $5.88 \times 10^{-16}$ \\
			\textsc{Rfl12} & 12 & 50 & $2.35 \times 10^{-15}$ \\
		\end{tabular}
    \caption{Run name, maximum refinement level, the corresponding minimum cell size, and the sink threshold density for the runs in our resolution study.}
    \label{tab:resolution}
\end{table}

\begin{figure*}
	\includegraphics[width=1.0\textwidth]{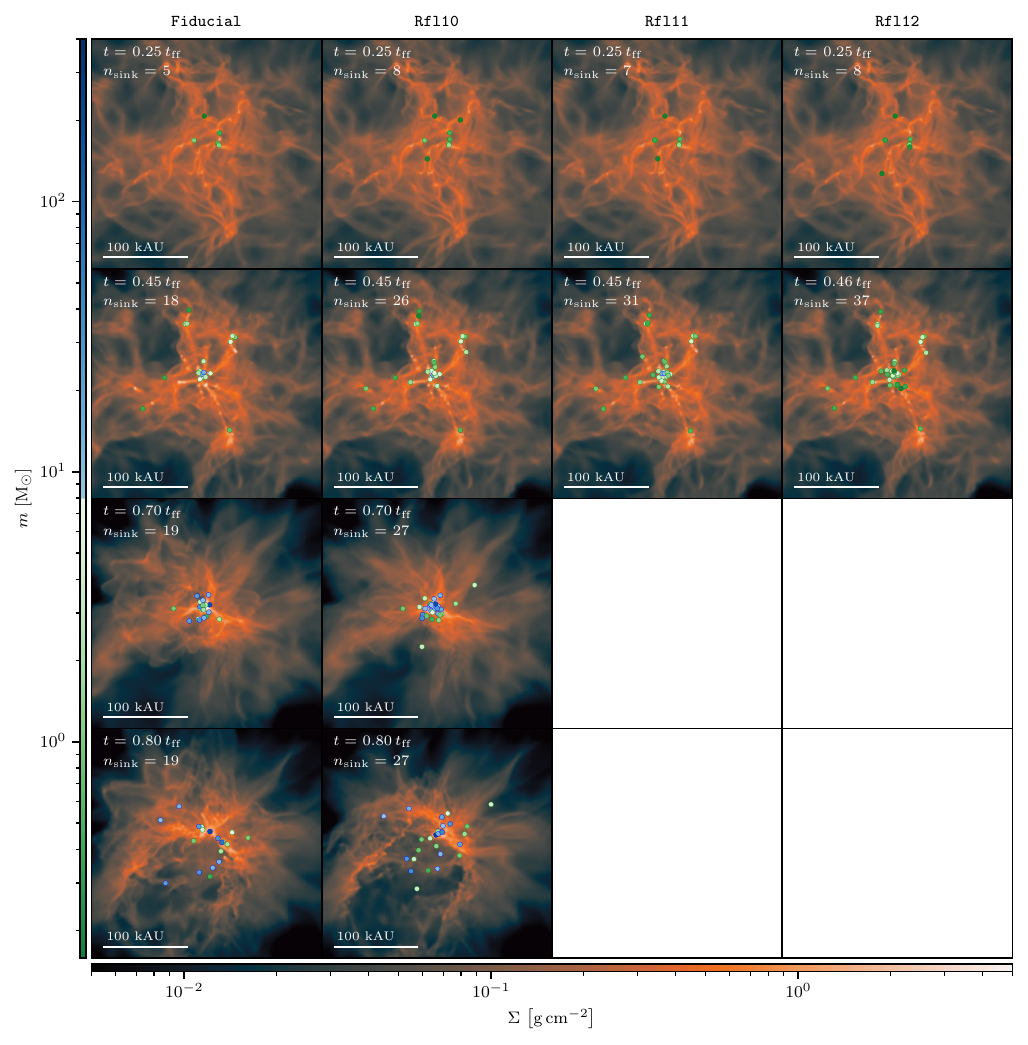}
    \caption{Evolution of the column density (from top to bottom) for different resolutions (from left to right). The higher the resolution, the more sink particles are created. The resolution has an impact on the evolution of the UC-HII region. Although the UC-HII region forms slightly earlier and appears to be more pronounced, the feedback is less efficient in regulating star formation with increasing resolution. Panels are blank for runs that were stopped at earlier times due to computational expense.}
    \label{fig:combinedResolution}
\end{figure*}

\begin{figure*}
      \centering
     \begin{subfigure}[b]{0.45\textwidth}
     \includegraphics[width=\columnwidth]{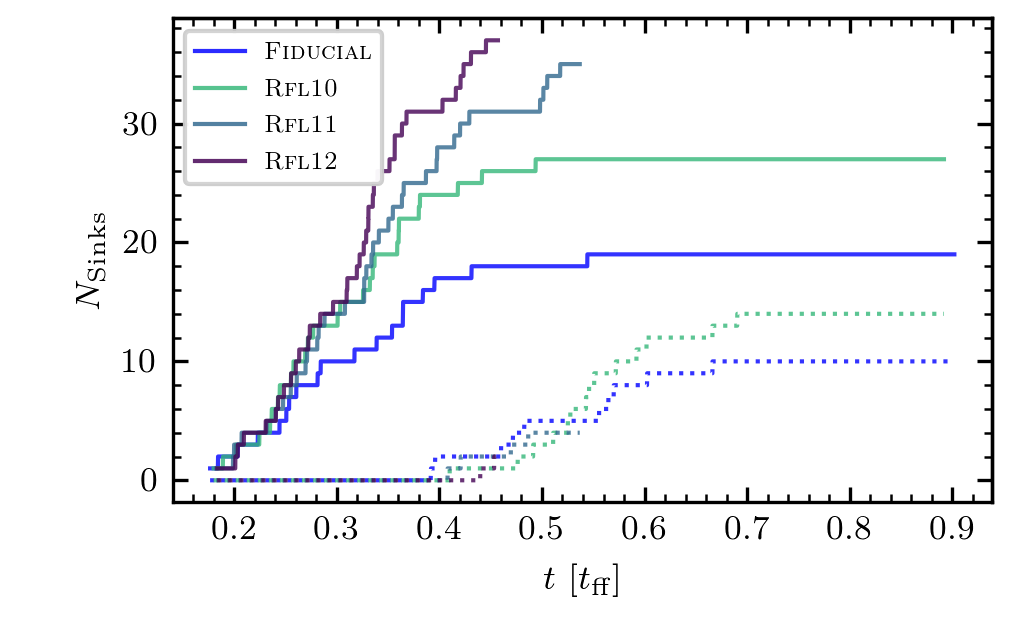}
     \end{subfigure}
    \hfil
     \begin{subfigure}[b]{0.45\textwidth}
    \includegraphics[width=\columnwidth]{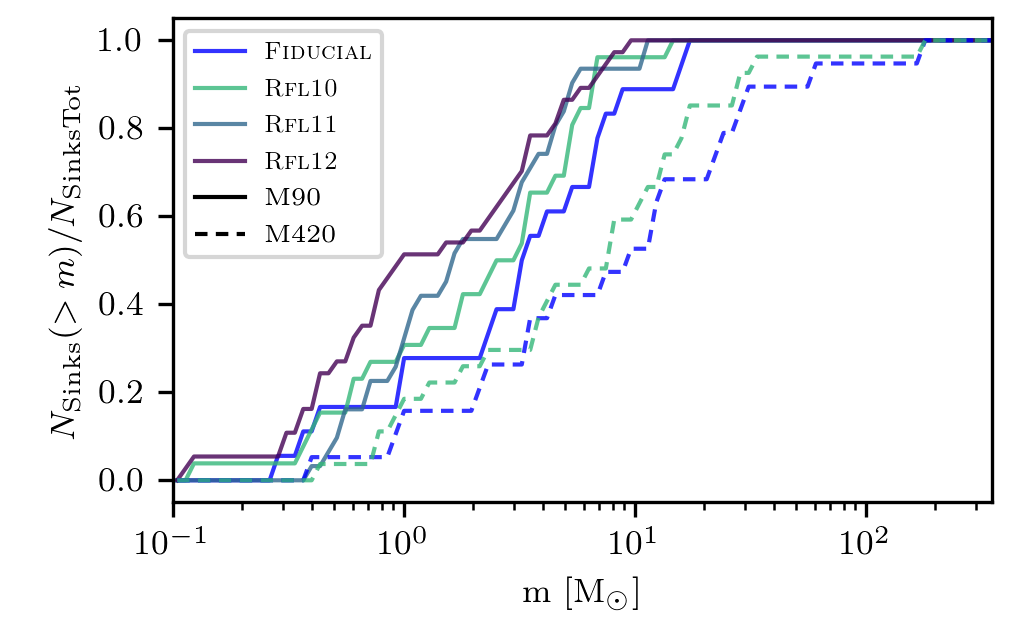}  
     \end{subfigure} 
     \caption{\texttt{Left}: Number of sink particles as a function of time (solid lines) and number of high mass sinks with $\geq 8M_\odot$ (dotted lines). \texttt{Right}: Cumulative sink mass distributions, normalized to the respective total number of sink particles. Two different times are shown: when a total mass of $90 \, \mathrm{M}_\odot$ has been accreted onto sinks (M90; solid lines) and when $420 \, \mathrm{M}_\odot$ (M420; dashed lines) have been accreted, respectively. M420 can only be presented for runs \textsc{Fiducial} and \textsc{Rfl10} since the higher-resolution runs had to be stopped earlier. Overall, runs with higher resolution show more fragmentation and the corresponding cumulative mass functions are slightly more bottom-heavy. For M420, there is no significant difference in the mass functions of \textsc{Fiducial} and \textsc{Rfl10}.} \label{fig:CMFresolution}
\end{figure*} 

\begin{figure}
     \centering
     \begin{subfigure}[b]{0.45\textwidth}
         \centering
         \includegraphics[width=\columnwidth,trim=0 0 0 0.056cm, clip]{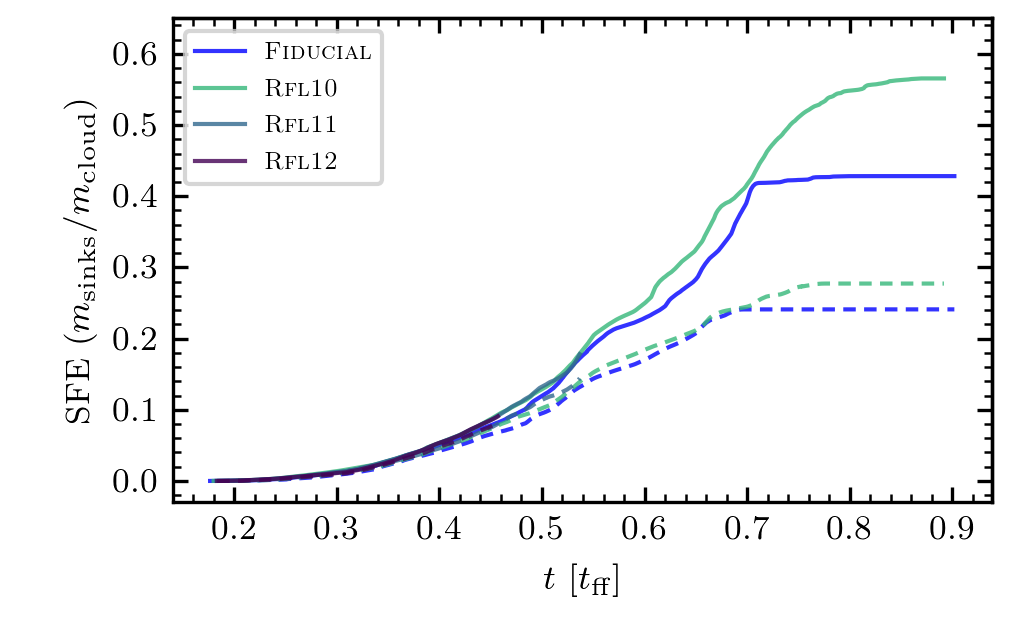}
     \end{subfigure}\\
     \begin{subfigure}[b]{0.45\textwidth}
         \centering
         \vspace{-1.03cm}
         \includegraphics[width=\columnwidth,trim=0 0 0 0.15cm, clip]{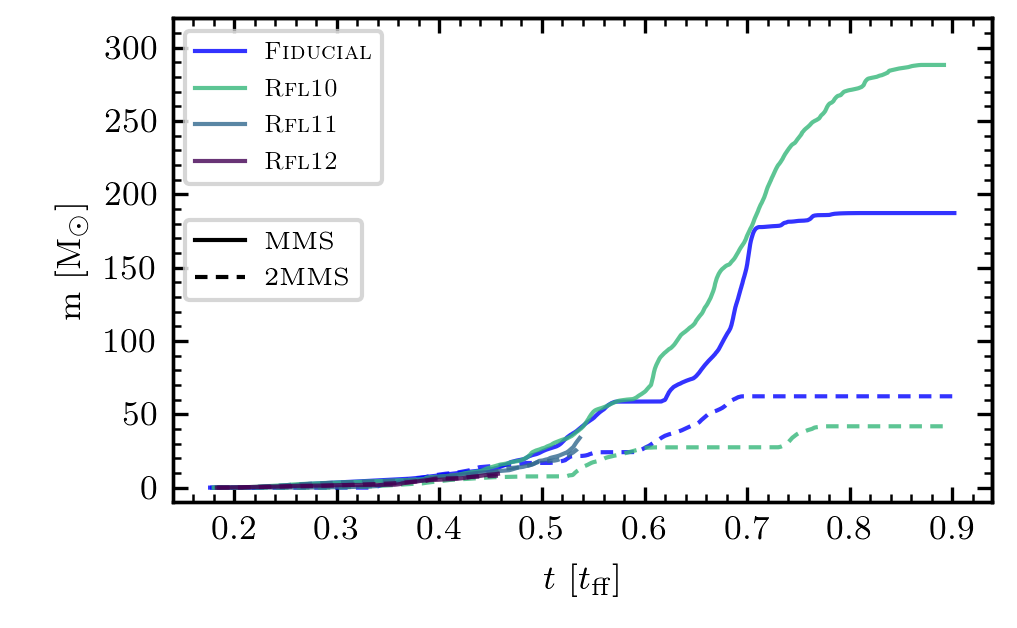}
     \end{subfigure}\\
     \begin{subfigure}[b]{0.45\textwidth}
         \centering
         \vspace{-1.04cm}
         \includegraphics[width=\columnwidth,trim=0 0 0 0.15cm, clip]{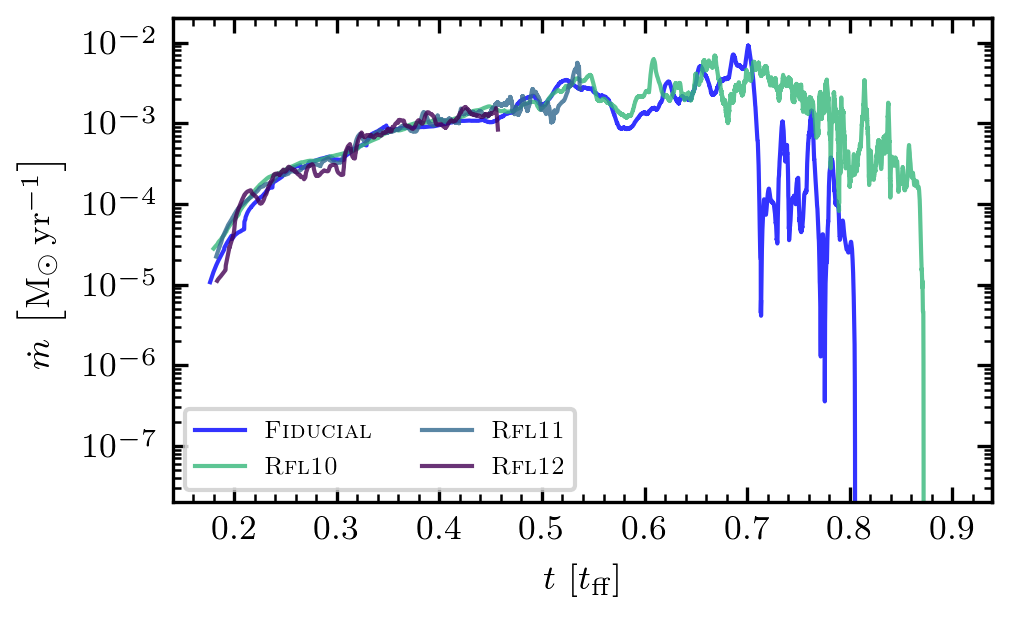}
     \end{subfigure}\\
     \caption{\texttt{Top}: Evolution of the star formation efficiency. Solid lines show the total SFE and dashed lines the total mass in companion stars (when the MMS is subtracted).
     \texttt{Centre}: Evolution of the mass of the most massive sink (MMS) and the second most massive sink (2MMS) in each simulation.
     \texttt{Bottom}: Total mass accretion rate onto sinks (smoothed). In the beginning all runs evolve in a similar manner. At later times, mass accretion continues for longer for higher resolution due to higher densities around the sinks. Thus, in run \textsc{Rfl10}, the SFE and the mass of the most massive sink particle grow beyond the ones found in the lower resolution run (\textsc{Fiducial}).}
     \label{fig:resolution}
\end{figure}

\begin{figure}
	\includegraphics[width=\columnwidth]{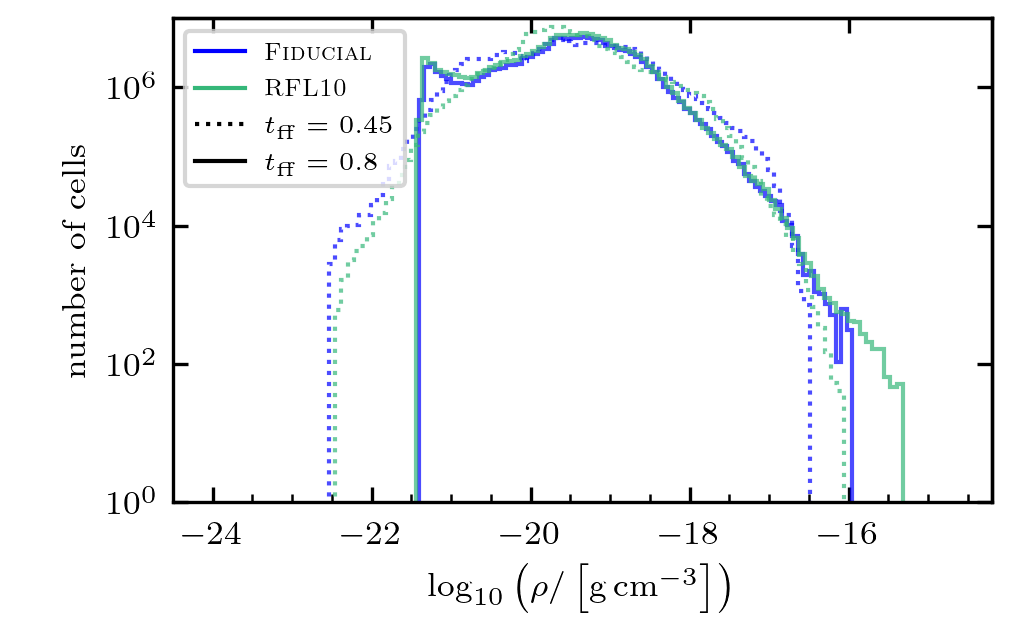}
    \caption{Density probability distribution functions (PDFs) within a sphere around the MMS with a radius of $10 \rm{kau}$. We compare runs \textsc{Fiducial} (blue) and \textsc{Rfl10} (green) at two different times, $0.45\,t_{\rm ff}$ (dotted lines) and $0.8\,t_{\rm ff}$ (solid lines), respectively. Higher densities are resolved in the higher resolution case.}
    \label{fig:resolutionDensity}
\end{figure}

\subsection{Resolution Study}
\label{Sec:ResolutionStudy}
We start from the \textsc{Fiducial} run with a maximum spatial resolution of $\Delta x =400$~au at refinement level 9 and increase the maximum refinement level in different simulations for the same initial conditions (see Tab. \ref{tab:resolution}). We increase the refinement level to 10 (run \textsc{RFL10}), 11 (run \textsc{RFL11}), and 12 (run \textsc{RFL12}), which corresponds to an effective resolution of $\Delta x =200$~au, $\Delta x =100$~au, $\Delta x =50$~au, respectively.  For all simulations, the minimum refinement level is set to 5. This corresponds to a $128^3$ base grid with a base resolution of $\sim 6,400 \, \mathrm{au}$ in the ambient medium.

\subsubsection{Morphological Evolution with Resolution}

Figure \ref{fig:combinedResolution} shows the time evolution of the column density of the four runs with different resolutions (the increase in resolution is shown from left to right). The simulations \textsc{RFL11} and \textsc{RFL12} had to be stopped much earlier than the lower resolution runs due to limitations in the computing time. 

The early evolution of the cloud (until $\sim 0.45 t_{\mathrm{ff}}$) is similar for all runs, modulo the fact that more sink particles are formed with higher resolution. For later times, comparing runs \textsc{RFL10} and \textsc{Fiducial}, the cavity blown by stellar feedback forms a bit earlier and seems to be more pronounced with higher resolution.

\subsubsection{Impact of Resolution on Sink Statistics}

The stars in our simulation are represented by sink particles.
Several criteria are checked before sink particles are formed or allowed to further accrete gas from within the sink accretion radius (see section \ref{sec:NumMeth}). 
One condition for the formation of sink particles is that the gas density within the accretion radius (2.5 cells on the highest refinement level) is greater than a certain density threshold, $\rho_{\rm thresh}$ (see eq. \ref{eq:rhothresh}). It depends on the smallest resolvable Jeans length at the maximum refinement level. Table~\ref{tab:resolution} lists the refinement level, the corresponding minimum cell size, and the sink threshold density for our resolution study.

During the initial collapse phase, the gas density is optically thin to infrared radiation and the core cools efficiently via dust thermal emission and behaves isothermally. When the gas density reaches $\rho\gtrsim 10^{-13}{\rm g}\,{\rm cm}^{-3}$, it becomes optically thick to infrared radiation. The core cannot cool efficiently anymore and behaves adiabatically \citep{Omukai2005}.
Therefore, for densities $>10^{-13}{\rm g}\,{\rm cm}^{-3}$, the Jeans mass increases with increasing density and no more fragmentation should occur. Hence, only at such high densities would a sink particle represent a single star. 
In other words, we would only expect convergence with increasing resolution if a Jeans density of $\sim 10^{-13}{\rm g}\,{\rm cm}^{-3}$ could be resolved \citep[e.g.,][]{Hubber2013}. In our setup, this would require a refinement level of 15. However, when sink particles are formed at lower densities, they may represent binaries or even higher-order multiples. 
Therefore, for any maximum resolution used in this work, 
sink particles could represent single stars, binaries, or higher-order systems. 

In the regime we resolve, it is expected that the number of sink particles, especially the number of lower mass sink particles, increases with higher spatial resolution. This is indeed the case (see Fig.~\ref{fig:CMFresolution}, left panel). Both, the number of low- and high-mass sink particles (sinks with more than $8\,\mathrm{M}_{\odot}$) increases (also see Fig. \ref{Fig.resolutionhist}). 
Additionally, the cumulative sink mass function (Fig.~\ref{fig:CMFresolution}, right panel) becomes slightly more bottom heavy with increasing resolution, at least at the time when $90\,{\rm M}_{\odot}$ have been accreted onto sinks (as shown by the solid lines). Later, the cumulative mass functions of the runs \textsc{Fiducial} and \textsc{RFL10} appear similar. However, overall, the lower resolution runs are lacking lower mass sink particles. 
We conclude that cloud fragmentation and sink formation are very sensitive to the numerical resolution in this regime, as expected. For a detailed analysis of the fragmentation process and the resulting sink mass distribution, it would be necessary to run simulations with higher resolution.

\subsubsection{Resolution Effects on Accretion and SFE}
As shown in Fig.~\ref{fig:resolution}, we find that all runs show a similar total mass accretion rate onto sinks until $\sim 0.7 \, t_{\mathrm{ff}}$ (see bottom panel). After this time, the total accretion rate of run \textsc{Fiducial} decreases significantly, while \textsc{RFL10} still accretes at a higher rate.
This results in a similar SFE at early times (Fig.~\ref{fig:resolution}, top panel); however, while the SFE of the \textsc{Fiducial} run stops at $0.42$ after $\sim 0.7 \, \mathrm{t_{ff}}$, the continued mass accretion in \textsc{RFL10} leads to a higher final SFE of $0.56$ (after $\sim 0.85 \, \mathrm{t_{ff}}$).
The SFE in companion stars, i.e., when subtracting the mass of the MMS from the total mass in sinks (dashed lines), we find that the SFE is similar for both \textsc{Fiducial} and \textsc{RFL10}, with values of $0.24$ and $0.27$, respectively. This means that the reason for the higher SFE in \textsc{RFL10} is the increase in the mass of the MMS. The MMS in \textsc{RFL10} ends up with $241 \, \mathrm{M}_\odot$, which is higher compared to the \textsc{Fiducial} run, where the most massive sink particle reaches $187.2 \, \mathrm{M}_\odot$ (see \ref{fig:resolution}, middle panel). However, the second most massive sink (2MMS) in both runs again have comparable masses, with the 2MMS of \textsc{Fiducial} being slightly more massive than the one formed in run \textsc{RFL10}. 

\subsubsection{UC-HII Expansion}

The MMS forms near the centre of the core and is surrounded by gas with high density as long as the sink particle is accreting. The density probability distribution function (PDF) within a sphere with a radius of $10 \, \rm{kau}$ around the MMS shows that \textsc{RFL10} generally resolves higher densities compared to the \textsc{Fiducial} run (see Fig. \ref{fig:resolutionDensity}). Due to higher densities in \textsc{RFL10} the recombination rate is higher, resulting in a decreased Str\"omgren radius and the UC-HII region is trapped for a longer period of time. During this time, the MMS is able to further accrete the mass and therefore becomes more massive. As soon as the feedback becomes strong enough to blow away the surrounding gas, the UC-HII region inflates, and further mass accretion is reduced and finally fully prevented. 

Nevertheless, the general evolution of the core collapse and dispersal is similar for all runs with different resolutions.
Therefore, $l_{\rm ref}=9$ is used to carry out a parameter study. At this resolution, it is possible to produce a statistically relevant sample of simulations, which will be later used as a basis for a comparison to the ALMAGAL observations.

\begin{figure*}
	\includegraphics[width=1.0\textwidth,trim=0.2cm 0 0.1cm 0cm]{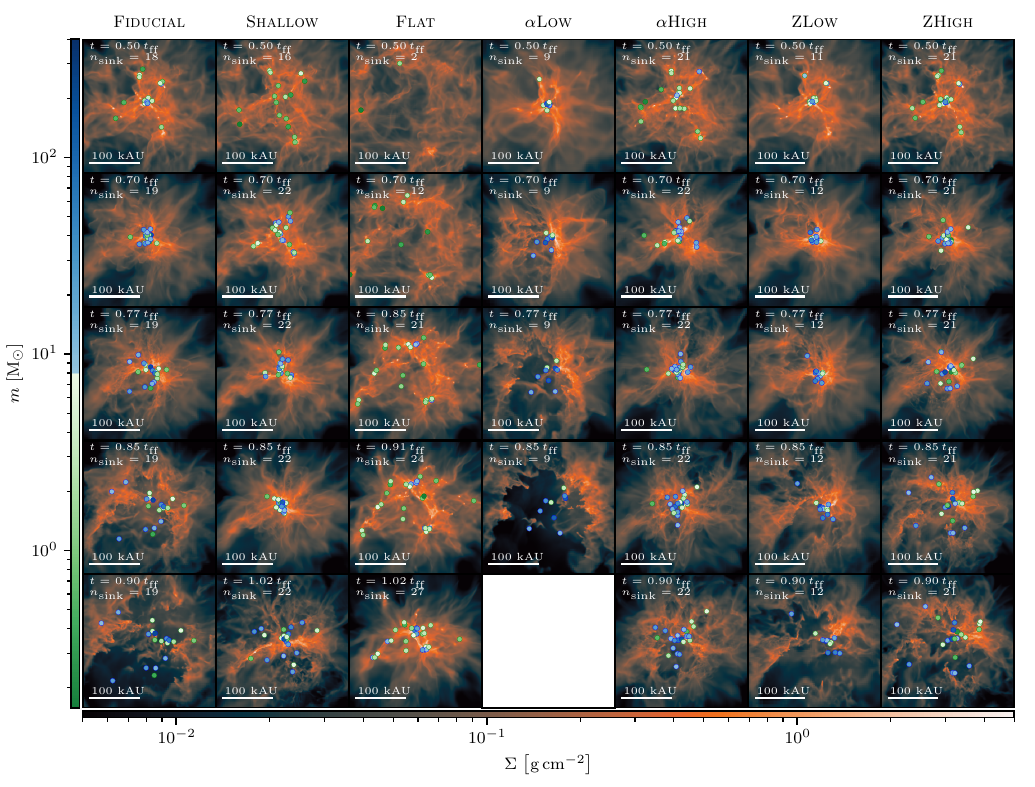}
    \caption{Time evolution (from top to bottom) of the column density distributions for different initial conditions according to tab. \ref{tab_parameter}. The flatter the density profile (columns one, two, three), the slower the collapse. A lower virial parameter (fourth column) leads to a faster collapse while a higher virial parameter (fifth column) leads to more sub-structures. Lower metallicity leads to less fragmentation (sixth column) compared to higher metallicity (last column), which promotes fragmentation. Because Run \textsc{$\alpha$Low} evolved more rapidly, the last panel is blank.}
    \label{fig:combinedParameter}
\end{figure*}

\section{Parameter Study}
\label{sec:ParameterStudy}

 \begin{figure*}
     \centering
     \begin{subfigure}[b]{0.45\textwidth}
     \includegraphics[width=\columnwidth]{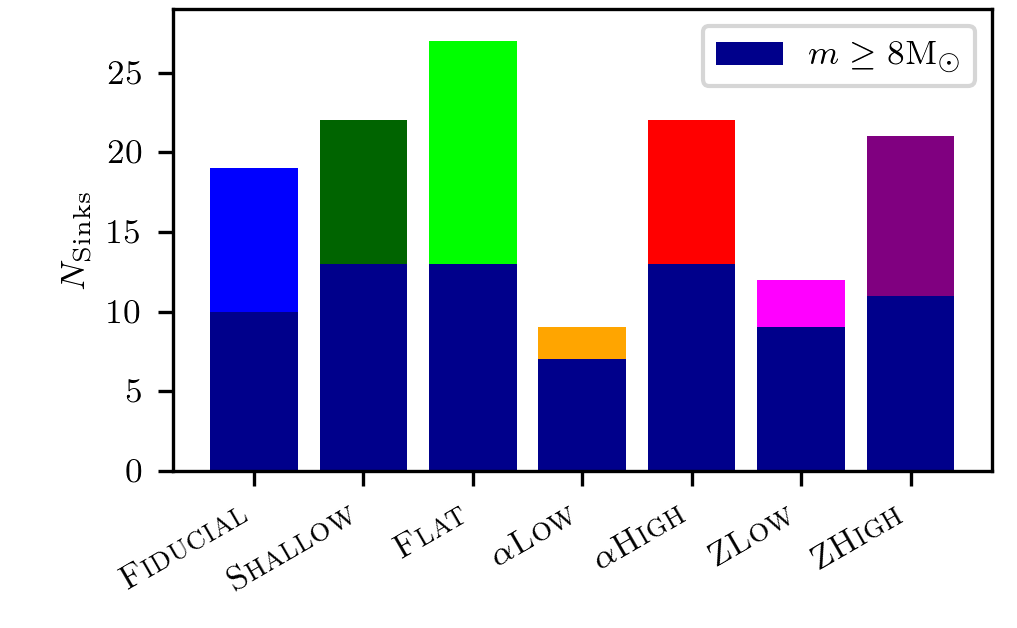}
     \end{subfigure}
     \hfil
     \begin{subfigure}[b]{0.45\textwidth}
     \includegraphics[width=\columnwidth]{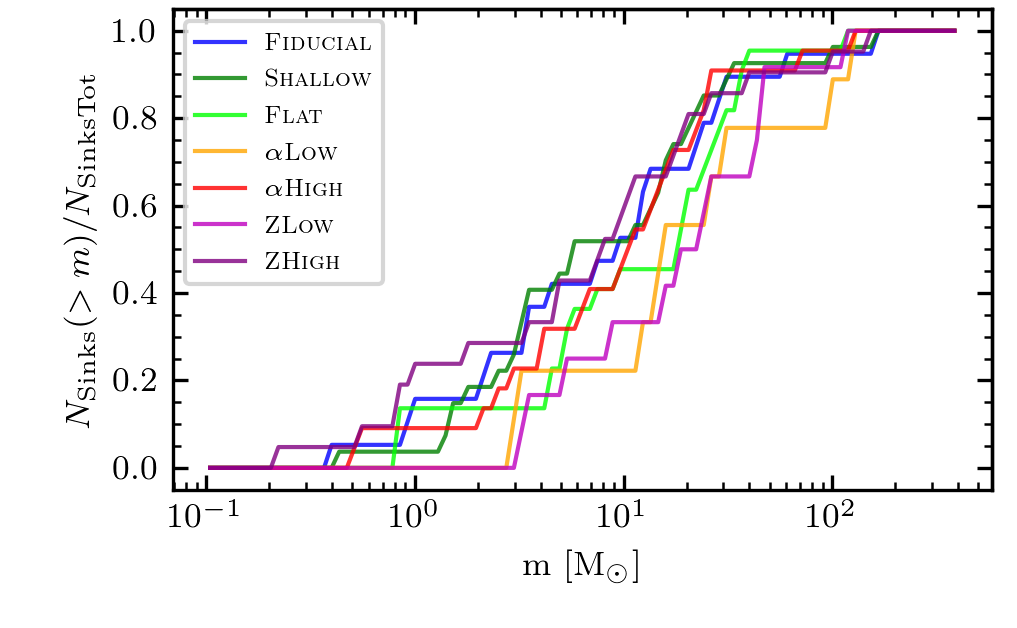}
     \centering     
     \end{subfigure}
     \caption{\texttt{Left}: Number of formed sink particles for the different runs. The dark blue parts indicate the number of high mass stars. \texttt{Right}: Cumulative sink mass distribution, normalized to the total number of formed sink particles of each run. Both plots represent the time when 95\% of the final mass has been accreted onto sink particles.  
     The flatter the initial density profile, the more fragmentation occurs. A low virial parameter as well as a low metallicity leads to fewer and very massive sink particles and the cumulative sink mass distribution is shifted towards top heavy. High virial parameter and metallicities have a minor impact on the fragmentation process and the cumulative sink mass distribution is comparable to the \textsc{Fiducial} run.
    }
     \label{fig:sinkbarparameter}
\end{figure*}

\begin{figure*}
     \centering
     \begin{subfigure}[b]{0.45\textwidth}
         \centering
         \includegraphics[width=\columnwidth,trim=0 0 0 0.056cm, clip]{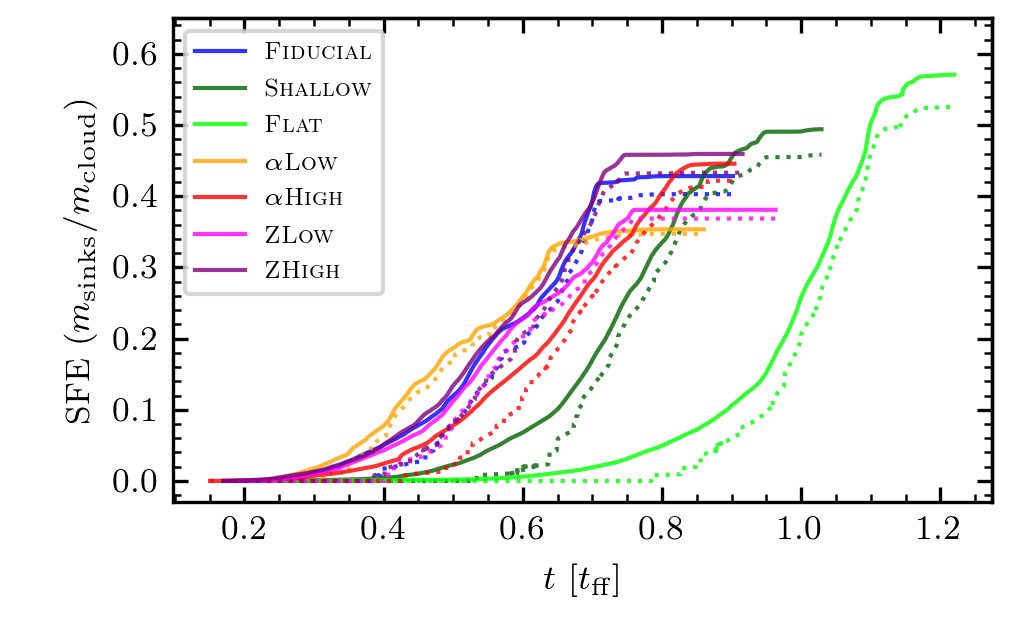}
     \end{subfigure}
     \hfil
     \begin{subfigure}[b]{0.45\textwidth}
         \centering
         \includegraphics[width=\columnwidth,trim=0 0 0 0.056cm, clip]{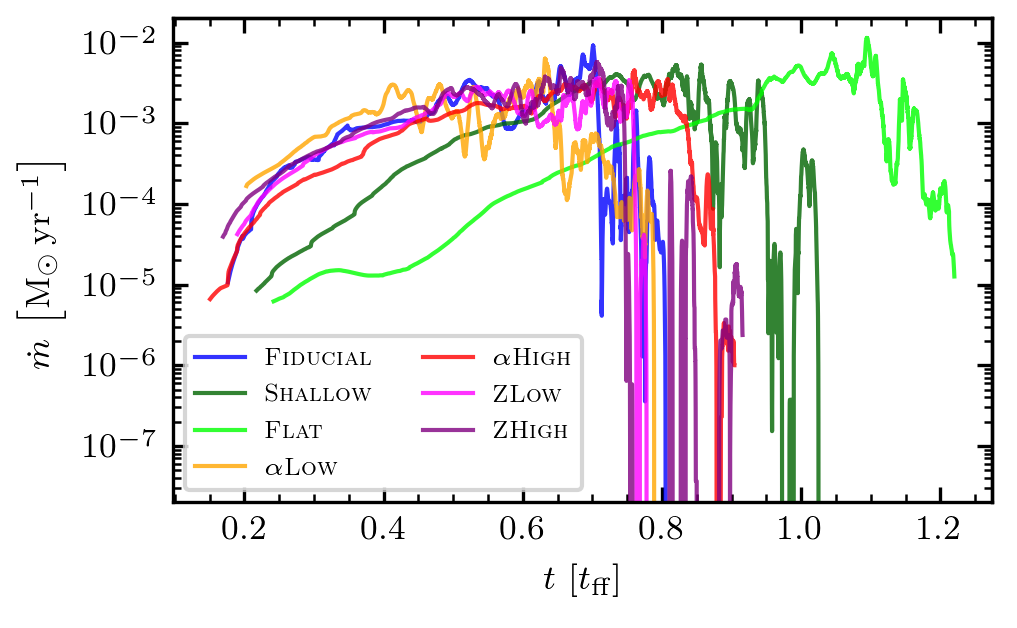}
     \end{subfigure}\\
     \begin{subfigure}[b]{0.45\textwidth}
         \centering
         \vspace{-1.03cm}
         \includegraphics[width=\columnwidth,trim=0 0 0 0.15cm, clip]{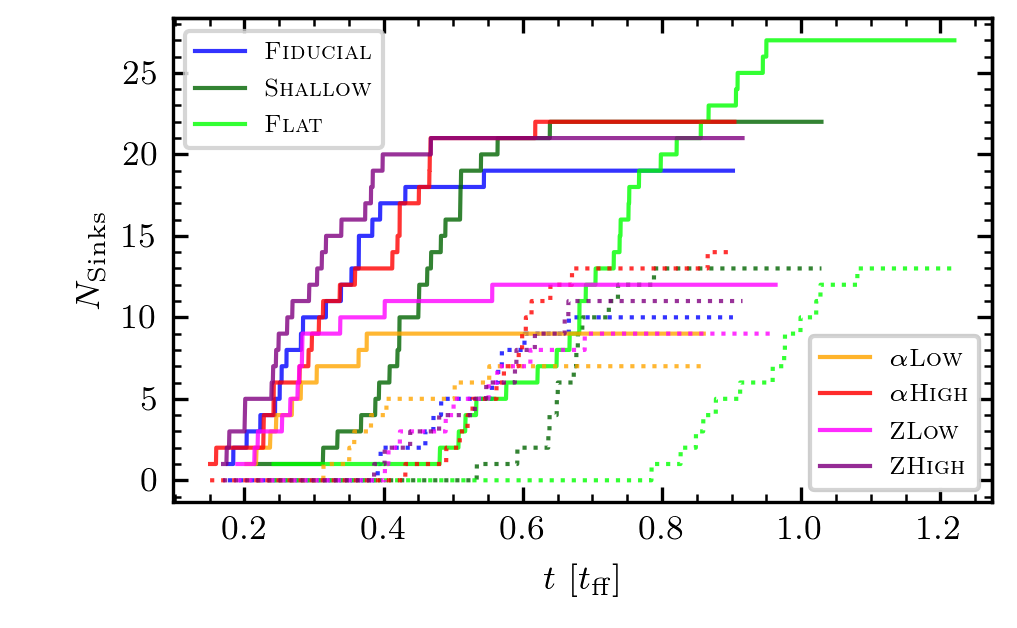}
     \end{subfigure}
     \hfil
     \begin{subfigure}[b]{0.45\textwidth}
         \centering
         \vspace{-1.03cm}
         \includegraphics[width=\columnwidth,trim=0 0 0 0.15cm, clip]{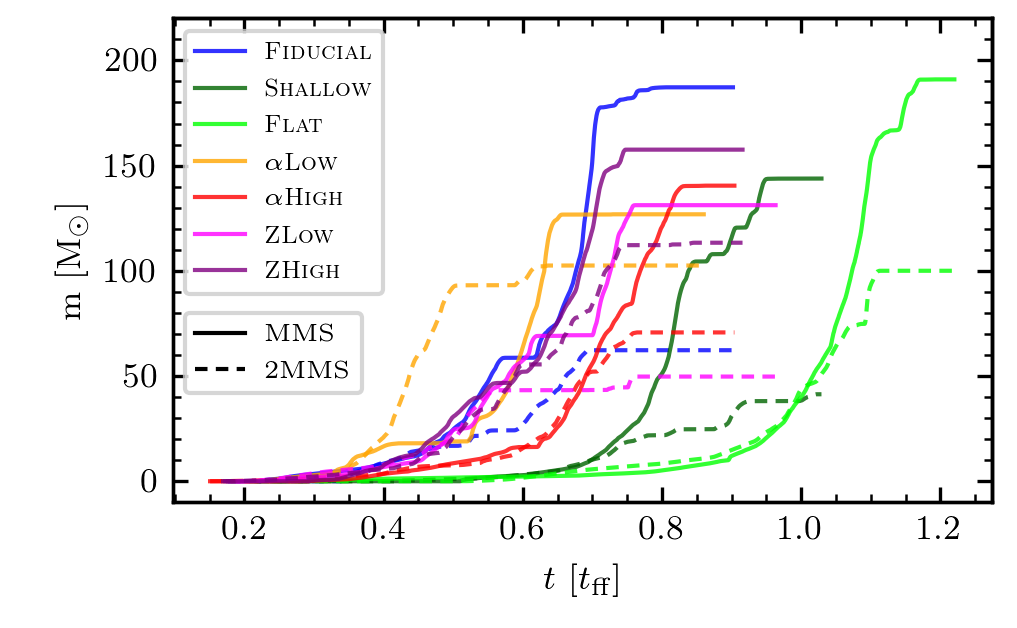}
     \end{subfigure}
     \caption{
     \texttt{Top Left}: Evolution of the star formation efficiency. The dotted lines represent the SFE of massive stars only.
     \texttt{Top Right}: Total accretion rate (smoothed). 
     \texttt{Bottom Left}: Number of sink particles, while the dotted line shows the number of high mass sinks with $\geq 8 \, \rm M_\odot$.
     \texttt{Bottom Right}: Mass of the most massive sink (MMS) and the second most massive sink particle (2MMS).
     Runs with flatter density profile collapse more slowly but in the end, they produce more fragments resulting in a higher SFE. The total accretion rate starts lower but accretion continues for longer. However, the mass of the MMS does not reach the one in the \textsc{Fiducial} run. Low virial parameter and low metallicities lead to fewer sink particles with reduced SFE and a reduced mass of the most massive sink. A high virial parameter and high velocity dispersion evolves similar to the \textsc{Fiducial} run, but the most massive sink particle is not as massive as in the \textsc{Fiducial} run.
   }
   \label{fig:parameter}
\end{figure*}

 \begin{figure*}
     \centering
     \begin{subfigure}[b]{0.45\textwidth}
     \includegraphics[width=\columnwidth]{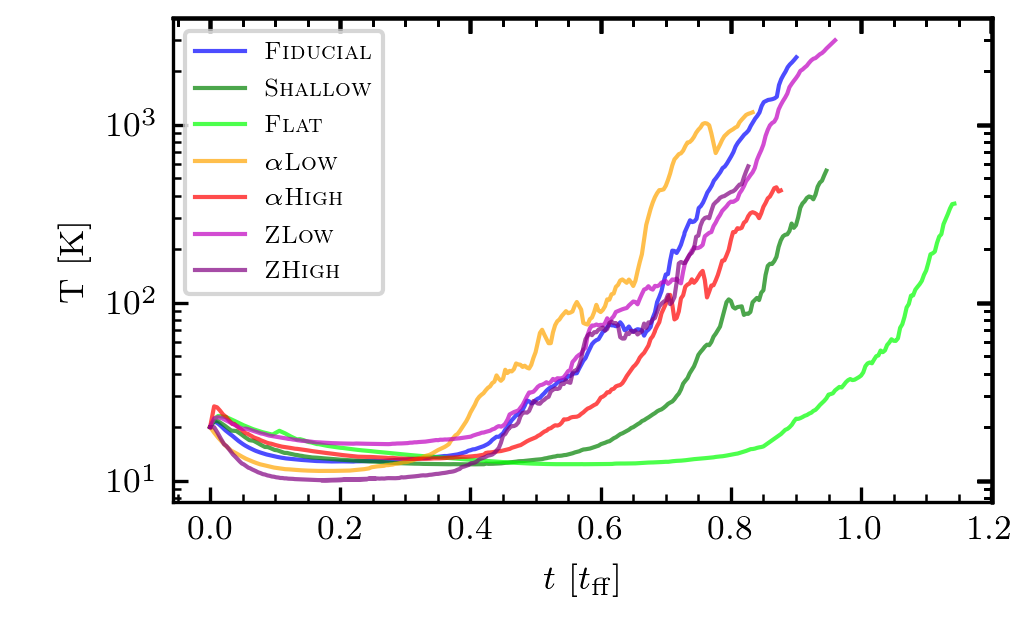}
     \end{subfigure}
     \hfil
     \begin{subfigure}[b]{0.45\textwidth}
     \includegraphics[width=\columnwidth]{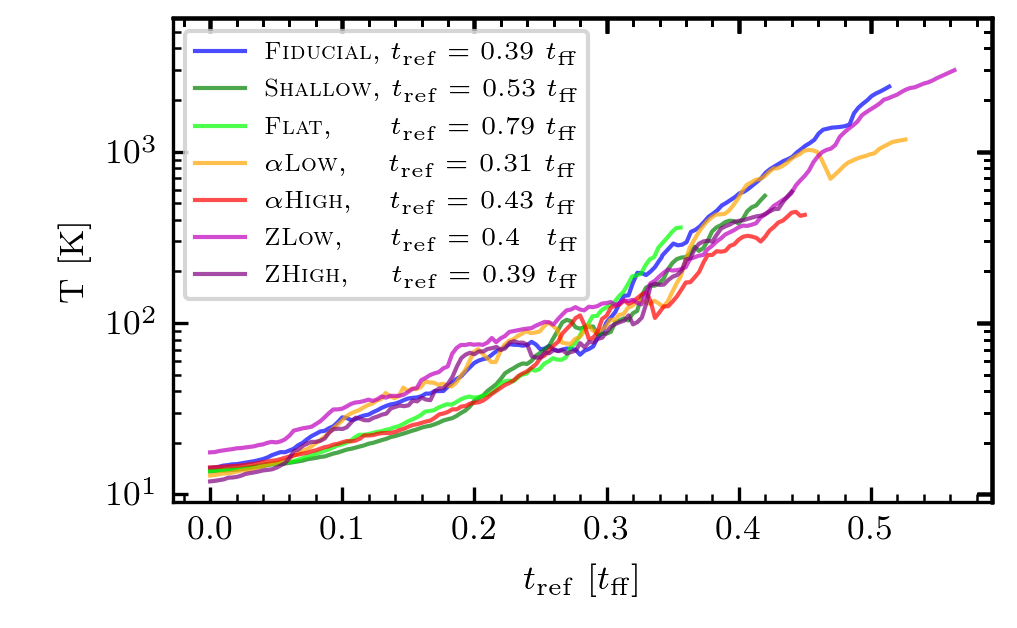}     
     \end{subfigure}
      \caption{\texttt{Left}: Mass-weighted temperature of the whole simulation domain. \texttt{Right:} Mass-weighted temperature, with $t_{\rm ref} = 0$ when first sink has $8 \, \rm M_\odot$. The runs with a delayed core collapse (\textsc{$\alpha$High}, \textsc{Shallow}, and \textsc{Flat}) also show a corresponding delay in the temperature evolution. Additionally, once a massive star has formed these cores heat up a bit more slowly but catch up with the other runs later on. The amount of metals affects the gas temperature due to more (\textsc{ZHigh}) or less (\textsc{ZLow}) dust cooling in the first $0.4\,t_{\rm ff}$.}
     \label{fig:MwgTparameter}
\end{figure*}

We perform a parameter study in which we vary the initial conditions (see sec. \ref{Sec. Initial Conditions}) of the collapsing core and investigate the resulting SFE and stellar feedback. Specifically, we compare simulations with different initial density profiles, virial parameters, and metallicities. The set of runs is summarized in Table \ref{tab_parameter}.

\subsection{Effects of the Initial Density Profiles}
The initial density profile of the cloud cores affects their evolution. The evolution of the column density for each run is shown in Fig. \ref{fig:combinedParameter}. A \textit{flatter} initial density profile (second and third column, runs \textsc{Shallow} and \textsc{Flat}) results in a slower core collapse. Therefore, turbulence has more time to shape the cloud before it collapses under gravity, and more substructures are formed, where sink particle formation takes place. The densities at the edges of the cloud with a \textit{flatter} density profile are higher compared to the \textsc{Fiducial} run, thus turbulence leads to additional sink formation in the outer regions. The cloud collapse is less centrally condensed. Therefore, the MMS in the run \textsc{Flat} is formed at a distance of around $100 \, \mathrm{kau}$ from the centre.

Flatter density profiles lead to more fragmentation and thus to an increasingly high number of sink particles (see Fig. \ref{fig:sinkbarparameter}, left panel). In both runs, \textsc{Shallow} and \textsc{Flat}, more high-mass stars are found, probably because feedback sets in later. Additionally, for a flat initial density profile (run \textsc{Flat}) also the number of lower- and intermediate-mass sink particles increases. Therefore, the normalized cumulative sink mass distribution for run \textsc{Flat} has, on the one hand, a dominant low mass end and, on the other hand, is also slightly shifted to be top heavy. 
Moreover, the slower core collapses of the runs \textsc{Shallow} and \textsc{Flat} lead to a significant time delay in sink formation (see Fig. \ref{fig:parameter}, bottom left panel). While the first sink particle in the \textsc{Fiducial} run is formed after $0.17 t_\mathrm{ff}$, the runs \textsc{Shallow} and \textsc{Flat} start with sink formation after $0.3 t_\mathrm{ff}$ and $0.43 t_\mathrm{ff}$, respectively (also see Fig. \ref{Fig.MassratioMMSparameter}).

The \textit{flatter} the initial density profile, the higher is the total SFE (see Fig. \ref{fig:parameter}, top left panel). Due to the slower core collapse, the initial mass accretion rate is lower compared to the \textsc{Fiducial} run (see Fig. \ref{fig:parameter}, top right panel). However, the slower collapse also extends the time period in which the sink particles can accrete mass before ionizing feedback stops the mass growth. This results in a final SFE of 0.57 in run \textsc{Flat} and 0.49 in run \textsc{Shallow}, which is higher than the SFE of the \textsc{Fiducial} run with 0.42 (see Fig. \ref{fig:parameter}, top left panel).

All three runs are dominated by one unique very massive sink particle. A flat initial density profile (run \textsc{Flat}) leads to a MMS with a mass of $191 \, {\rm M}_{\odot}$, which is even more massive than the MMS in the \textsc{Fiducial} run, while the MMS in run \textsc{Shallow} only reaches $144 \, {\rm M}_{\odot}$ (see Fig. \ref{fig:parameter}, bottom right).

The time-delayed core collapse affects the temperature evolution of the core, which starts rising when the first sink particle reaches $8 ~\rm M_\odot$ and starts to emit ionizing radiation, which heats the surrounding (see Fig. \ref{fig:MwgTparameter}, left panel).
At the reference time, $t_{\rm ref} = t - t_{\rm m}$, where $t_{\rm m}$ is the time when the first sink particle reaches $8 ~\rm M_\odot$, the temperature is similar (around $15 ~ \rm K$) for all three runs (see Fig. \ref{fig:MwgTparameter}, right panel). There is a clear trend that slowly collapsing runs also have a slower temperature rise compared to the \textsc{Fiducial} run but they catch up after $0.27 ~ t_{\rm ff}$.

\subsection{Effects of the Initial Virial Parameter}
The virial parameter gives the relation between the kinetic and gravitational energies of the initial core setup. A low virial parameter indicates that less turbulence disturbs the gravitational collapse, leading to the formation of fewer substructures during the collapse and vice versa (see Fig. \ref{fig:combinedParameter}).
For a low virial parameter (run \textsc{$\alpha$Low} with $\alpha_{\rm vir}=0.2$), the gravitational energy is dominant and the collapse is efficiently feeding the central hub where almost all sink particles are formed. The core is able to collapse faster than in run \textsc{Fiducial}. 
In run \textsc{$\alpha$High} with $\alpha_{\rm vir}=1.2$, the kinetic energy dominates and the cloud core collapses more slowly than in the \textsc{Fiducial} run.

A low turbulent energy (run \textsc{$\alpha$Low}) leads to a reduction of fragmentation and sink particle formation resulting in nine sink particles only, while the \textsc{Fiducial} run consists of 19 sinks (see Fig . \ref{fig:sinkbarparameter}, left panel). A high velocity dispersion (run \textsc{$\alpha$High}) leads to more fragmentation and 22 sink particles are formed.
The formed sink particles in run \textsc{$\alpha$Low} are very massive and thus the cumulative sink mass distribution is more top heavy (see Fig . \ref{fig:sinkbarparameter}, right panel). On the other hand, the shape of the mass distribution of \textsc{$\alpha$High} and the \textsc{Fiducial} run is similar.

The SFE in run \textsc{$\alpha$High} increases slower than in the \textsc{Fiducial} run, yet overall the star formation is only time-delayed and the SFE reaches a value of $0.4$, which is slightly higher than in \textsc{Fiducial} (see Fig. \ref{fig:parameter}, top left panel).
The final SFE of run \textsc{$\alpha$Low} is reduced to $0.35$. The ionizing photon emissivity from the few but very massive sinks increases faster compared to the \textsc{Fiducial} run and quenches mass accretion earlier (see Fig.~\ref{fig:parameter}, top right panel and Fig. \ref{Fig.parameter_total_LYC}). 
The increased turbulence in run \textsc{$\alpha$High} delays mass accretion onto (massive) sinks, leading to a lower ionizing photon rate (see Fig. \ref{fig:parameter}, top right panel and Fig.  \ref{Fig.parameter_total_LYC}). As a result, cloud disruption occurs later, extending the period during which sink particles can accrete mass, allowing the SFE to eventually catch up with the \textsc{Fiducial} run (see Fig. \ref{fig:parameter}, top left panel).

Compared to the MMS in the \textsc{Fiducial} run ($187 \,  \rm M_\odot$), the MMS has a reduced mass for both runs \textsc{$\alpha$Low} and \textsc{$\alpha$High} with final masses of $127 \, \rm M_\odot$ and $140 \, \rm M_\odot$, respectively (see Fig \ref{fig:parameter}, bottom right panel). With a low virial parameter, gravity is more important and thus, the accretion rate of the MMS increases compared to the \textsc{Fiducial} run (see Fig. \ref{fig:parameter}, bottom left panel). However, the earlier stop of mass accretion in run \textsc{$\alpha$Low} due to ionizing radiation feedback from the very massive sinks limits the MMS to grow more massive than the MMS in the \textsc{Fiducial} run. On the other hand, in run \textsc{$\alpha$High} the higher amount of turbulence hinders mass accretion onto the MMS.

For run \textsc{$\alpha$High}, the numbers of low and intermediate mass sinks are similar to the \textsc{Fiducial} run, but in total more massive sink particles (14 in total) are formed (see Fig. \ref{fig:parameter}, bottom left panel). Thus, \textsc{$\alpha$High} ends up with more high mass sinks which, individually, are not as massive as in the \textsc{Fiducial} run. In run \textsc{$\alpha$Low}, seven out of nine sink particles end up being massive.

The faster collapse in run \textsc{$\alpha$Low} results in a faster mass-weighted temperature increase, but the evolution is similar to the \textsc{Fiducial} run (see Fig. \ref{fig:MwgTparameter}, left panel).
The slower collapse and slower star formation in run \textsc{$\alpha$High} is also reflected in the temperature evolution: when considering the increase in temperature after $t_{\mathrm{ref}}$, run \textsc{$\alpha$High} is similar to the two runs with initially flatter density distributions (see Fig. \ref{fig:MwgTparameter}, right panel).

\subsection{Effects of Metallicity}
The gas and dust metallicity has an impact on the evolution of (massive) clouds and cores. 
Metal lines and dust cool the gas and their absence may result in a suppression of fragmentation in lower metallicity environments \citep{Omukai2005}. In addition, dust grains shield ionizing photons in high-density filaments \citep[e.g.][]{2020Fukushima}. At low metallicities, ionizing photons can penetrate further into higher-density gas where the star formation occurs and aid to disrupt the parental cloud. The temperature evolution of prestellar clouds changes with different metallicities \citep{Omukai2005}.

Starting with a lower initial metallicity, run \textsc{ZLow} shows less fragmentation and sink formation (see Fig. \ref{fig:combinedParameter}) than the \textsc{Fiducial} run. Most sink particles are formed in the central region and the collapse is faster. Only 12 sink particles are formed out of which nine become massive (see Fig. \ref{fig:sinkbarparameter}, left panel). Therefore, the cumulative sink mass distribution of run \textsc{ZLow} is shifted towards top heavy (see Fig. \ref{fig:sinkbarparameter}, right panel).
However, the cloud with a higher initial metallicity (run \textsc{ZHigh}) evolves similar to the \textsc{Fiducial} run. In total, 21 sink particles are formed, while 11 grow massive. This is slightly higher compared to the \textsc{Fiducial} run with 19 sink particles and 10 massive ones. However, the mass distribution of run \textsc{ZHigh} is comparable to the one of the \textsc{Fiducial} run.

During the first $0.25 \,{\rm t_{ff}}$ the formation of sink particles of the run \textsc{ZLow} is similar to runs \textsc{ZHigh} and \textsc{Fiducial}. After this time, fragmentation in \textsc{ZLow} is more and more suppressed, while in run \textsc{ZHigh} sink particle formation is accelerated (see Fig. \ref{fig:parameter}, bottom left panel). The lower number of sink particles in run \textsc{ZLow} leads to a reduction in the total SFE to 0.38, 
while run \textsc{ZHigh} reaches $\sim$0.46,
which is even higher than in the \textsc{Fiducial} run (see Fig. \ref{fig:parameter}, top left panel).
The accretion rates of all three runs are initially similar (see Fig. \ref{fig:parameter}, top right panel). However, mass accretion in run \textsc{ZHigh} lasts longer compared to the \textsc{Fiducial} run, causing the overall higher SFE. 
Nevertheless, the mass of the MMS ($158 \, \rm M_\odot$) is still lower in run \textsc{ZHigh} than in the \textsc{Fiducial} run ($187 \, \rm M_\odot$). The MMS in run \textsc{ZLow} reaches only $131 \, \rm M_\odot$.

The metallicity has an impact on the cooling of gas via dust. Higher metallicity leads to more efficient cooling and vice versa. The simulation starts with a gas temperature of $20 ~\rm K$, which is initially slightly cooled. During the first $0.4 \, t_\mathrm{ff}$ the mass-weighted temperature of run \textsc{ZHigh} is cooler ($\sim 10 ~\rm K$) than that of the \textsc{Fiducial} run with $\sim 13 ~\rm K$ and run \textsc{ZLow} with $\sim 16 ~\rm K$ (see Fig. \ref{fig:MwgTparameter}, left panel). 
After $\sim 0.4 \, t_\mathrm{ff}$, when the sinks become massive, the temperature increases in all runs. However, when considering the time evolution as normalized to $t_{\mathrm{ref}}$, \textsc{ZLow} still has a slightly increased temperature with respect to \textsc{ZHigh} and \textsc{Fiducial} until $t_{\mathrm{ref}} \sim 0.32\,t_{\mathrm{ff}}$ (see Fig. \ref{fig:MwgTparameter}, right panel).

\section{Discussion}
\label{sec:Discussion} 

\subsection{Effects of the Initial Density Distribution}
If the density distribution follows a power law ($\rho \sim r^{-1.5}$ or $\rho \sim r^{-2}$) the gravitational collapse leads to one central massive star. This is in agreement with previous work \citep{Girichidis2011,Rosen2019, 2023Kleptiko}.
However, even with an initially uniform density distribution, only one very massive sink particle (with mass $M_{\rm MMS} \sim 2 \cdot M_{\rm 2MMS}$) is formed, which dominates the gravitational potential. 
Consistent with our simulations, \cite{Girichidis2011} find that an initially uniform density leads to sink formation in filaments in the outer regions. However, in \cite{Girichidis2011}, the first sink particle that forms always evolves into the most massive one. In our simulations, the MMS is not necessarily the first formed sink particle (see Fig. \ref{Fig.MassratioMMSparameter}). This might be the result of stellar feedback, which is not included in the simulations of \cite{Girichidis2011}. 

\subsection{Effects the Initial Virial Parameter}
\cite{2017Dale} find a decreasing SFE for cores with an increasing virial parameter (from $\alpha_{\mathrm{vir}}=$ 0.7 to 1.5) simulating a larger-scaler setup than ours of a 10$^4 \, M_\odot$ cloud core within a radius of 5 pc. Also, \cite{2005KrumholzMckee} reports that the star formation rate decreases with an increasing virial parameter. However, in our study the overall SFE increases with the virial parameter (see Fig.~\ref{fig:parameter}, top left), which is again a result of stellar feedback. 
The subvirial core undergoes a fast global collapse with high accretion rates and photon emissivity, in agreement with \cite{Rosen2019}\footnote{Note that the simulations of \citet{Rosen2019} study cores with a similar gas surface density; however, the cores are smaller and less massive, and therefore the simulations afford a higher effective spatial resolution of 20~au.} and \cite{2005KrumholzMckee}.
The resulting higher source luminosities (intrinsic plus accretion) lead to high radiative heating rates and an efficient warm-up of the core (see left panel of Fig.~\ref{fig:MwgTparameter}), which suppresses fragmentation. However, high accretion rates result in more massive sinks and their high photon emissivities destroy the cloud faster causing lower SFEs. On the other hand, a higher level of turbulence leads to more turbulent fragmentation (also seen by \cite{Rosen2019}). As a result, a higher number of massive sinks is formed but they are less massive on average. The lower-mass MMS leads to less radiative heating, and the core is cooler than in the \textsc{Fiducial} run or in the low $\alpha_{\mathrm{vir}}$ case. Additionally less massive sinks have a lower photon emissivity, thus star formation proceeds for a longer time.

\subsection{Effects of Metallicity }
Metals cool the gas, which favours fragmentation \citep{Omukai2005}. In particular, dust cooling promotes the fragmentation within star forming clumps and cores. Simulations of star formation in the early universe show a tendency to result in a top-heavy IMF at lower metallicities ($10^{-4} \, Z_\odot$ and $10^{-5} \, Z_\odot$) \citep{2008Clark, 2011Dopcke, 2013Whalen, Latif2022}. However, \cite{2009bate,2019Bate} suggest that at near-solar metallicities, the overall properties of stars, including the IMF, are relatively insensitive with metallicity, supporting the idea that the IMF is largely universal.
Previous studies \citep{2014Tanaka,2022Matsukoba} show that the fragmentation of accretion discs is enhanced in the metallicity rage of $10^{-2} \, Z_\odot \, - \, 10^{-5} \, Z_\odot$. However, in this study, we do not resolve disc fragmentation and also use higher metallicities ($0.5 \, Z_\odot \, - \, 2 \, Z_\odot$).
For higher metallicities radiation pressure might become more important \citep{2020Fukushima} and therefore, pushing gas outwards.
\cite{2021Chen} reports that metallicities greater than $0.1 \, Z_\odot$ may limit the mass accretion of massive stars through the formation of low-mass companions.
Overall, the SFE tends to increase with higher metallicity, which is shown in \citep{2020Fukushima,2020He,2021Ali}, and can also be seen in our results.

\subsection{Trends in Star Formation Efficiency}
The overall SFE in our simulations is 35\% to 57\% depending on the initial conditions. The lowest SFE (35\%) is reached in run \textsc{$\alpha$Low} with a lower virial parameter compared to the \textsc{Fiducial} run. We see a clear increase in SFE (from 0.42 to 0.57) with an increasing central gas density surface from the \textsc{Fiducial} run with $\Sigma = 0.1 $ g cm$^{-2}$ ($\hat{=}~ 480 \, \mathrm{M}_\odot ~ \mathrm{pc}^{-2}$) to run \textsc{Flat} with $\Sigma = 0.39$ g cm$^{-2}$ ($\hat{=}~ 1800 \, \mathrm{M}_\odot ~ \mathrm{pc}^{-2}$). Therefore, a more rapid collapse does not necessarily lead to a higher SFE. 

Our SFE is perhaps slightly higher than expected (see discussion of code limitations on the SFE below). 
However, previous results show a strong dependence of the SFE on the core surface density and the density profile. For example, in low surface density clouds with $\Sigma= 0.01 \, \mathrm{g~cm}^{-2}$ ($\hat{=}~ 48 \, \mathrm{M}_\odot ~ \mathrm{pc}^{-2}$), \cite{2018Geen} find that the SFE can range from 6 to 23\% depending on the initial conditions. 
Higher SFEs (up to 21 \% and 32\%) are found in clouds with 
$\Sigma \sim 2.1 \, {\rm g ~cm}^{-2}$ ($\hat{=}~ 10^4 \, \mathrm{M}_\odot ~\mathrm{pc}^{-2}$)\footnote{The column density threshold for massive star formation suggested by \cite{2008Krumholz} is 1 g cm$^{-2}$ or $ 4.7 \times 10^3 \, \mathrm{M}_\odot ~ \mathrm{pc}^{-2}$.} \citep{2017Howard, 2018Grudic} and steep density profiles \citep[][SFE  $\geq $50\%]{2019Rahner}. For an initial gas surface density of 0.28 g cm$^{-2}$ ($\hat{=}~ 1300 \, \mathrm{M}_\odot ~ \mathrm{pc}^{-2}$), \cite{2018Kim} report a SFE of 51\% while the SFE in \cite{2016Raskutti} reaches over 60 \% with only 0.05 g cm$^{-2}$ ($\hat{=}~ 250 \, \mathrm{M}_\odot ~ \mathrm{pc}^{-2}$).
\cite{2021Fukushima} report that if the surface density reaches 0.02 g cm$^{-2}$ ($\hat{=}~ 100 \, \mathrm{M}_\odot ~ \mathrm{pc}^{-2}$) ionizing feedback becomes increasingly inefficient in regulating mass accretion due to the deep gravitational potential well and thus the SFE increases.

\subsection{Feeback Effects}
\subsubsection{Ionizing Radiation}
Stellar feedback is known to reduce the SFE. Ionizing radiation is a main feedback mechanism, which limits the final SFE and is crucial to model HII regions \citep{2015Geen,2017Dale, 2018Haid}. We use the approximation of a black body spectrum to model the ionizing photon fraction without considering the stellar atmosphere, which might lower the ionizing photon rate \citep{1979Kurucz,2020Kuiper}. The SFE is very sensitive to the ionizing photon emissivity \citep{2022Dobbs}, which we also see in our simulations. A fast increase in the ionizing photon rate results in a short mass accretion period and thus lower SFE. \cite{OlivierG2021} shows that it might be useful to take into account UV attenuation by dust in HII regions to not overestimate the ionization rate. 
Furthermore, FUV radiation which can photo-dissociate H$_2$ and CO and cause photoelectric heating is not included yet and can suppress star formation \citep{2020Inoguchi,2022Fukushima}. However, in a follow-up code development paper of FLASH \citep{2025Rathjen} a self-consistent adaptive ISRF is implemented. They find that, compared to photoionization, the IRSF is not that important for regulating star formation on scales of several hundred parsec.

\subsubsection{Radiation Pressure}
Radiation pressure is important to include in numerical simulations on different scales even if it is not the dominant feedback mechanism \citep{Peters2017,Rosen2020,2023Kleptiko}. Especially the interplay of photo-ionization and RP seems to be important at later stages and may limit the final mass of high-mass stars \citep{2018Kuiper}. In addition, RP supports the growth of the HII region \citep{2018Ali,2021Ali}. 
RP from non-ionizing radiation, which is not included in our simulations, might become increasingly important at higher resolutions in higher density regions \citep{2018Kuiper}. Neglecting the acceleration due to non-ionizing radiation pressure could lead to an underestimation of feedback effects during the early stages, potentially allowing somewhat higher accretion rates and final stellar masses than would be expected if this mechanism were included.

\cite{2018Kuiper} report that on smaller scales the radiation feedback caused by dust thermal emission is most effective in suppressing mass accretion onto a massive protostar. However, \cite{2018Kuiper} are resolving mass accretion through a disc onto a protostar because of the use of a spherical grid. They also use a fixed dust opacity of $\kappa_{\rm dust} = 4 \times 10^4 \mathrm{cm}^2 \mathrm{g}^{-1}$ in the FUV regime, which is much higher than our typical dust opacity, which depends on the dust temperature and is calculated self-consistently (see equation \ref{eq:dustOpacity} and Fig.~\ref{fig:fiducialrunTemp}; typically we have Planck mean opacities between 10 - 10$^3 \, \mathrm{cm}^2 \mathrm{g}^{-1}$). 
With the resolution of our \textsc{Fiducial} run ($\sim 400 ~\rm au$) we are not able to resolve the accretion disc around a massive star (see also Sec. ~\ref{Sec:ResolutionStudy}). Resolving the accretion disc around the MMS might limit its final mass and thus, also its radiative feedback.
With a resolution of $\sim 20 ~\rm au$,  \cite{2023Kleptiko} and \cite{Rosen2020} also conclude that RP alone is not able to suppress mass accretion. 

Looking at GMC scales, \cite{2018Kim} (fiducial model with
$M = 10^5 M_\odot$ and $R = 20 \rm pc$) says that ionizing feedback has a greater impact on cloud destruction than RP and only considering RP leads to a higher SFE, which is in agreement with our results. The larger the scales considered, the lower the average density, and the less impactful is the feedback by RP on dust \citep{2018Krumholz}. 
However, they also emphasize that the effects of both feedback mechanisms are not linear and act in concert. By calculating direct RP from a star, resolving the dust sublimation front (between 100 - 1000 au) can play an important role \citep{2010bkuiper,2018Krumholz} in not overestimating the escape fraction of ionizing radiation and thus the efficiency of RP. In our simulation, according to \cite{2018Krumholz}, we only resolve the dust sublimation radius for the very massive sinks ($>40\, \mathrm{M}_\odot$). However, we also see that neglecting RP does not change the overall SFE. Thus, we speculate that reducing the efficiency of RP by using a subrid model that describes the effects of dust sublimation would not strongly affect the SFE.

In addition to RP, radiative heating by non-ionizing photons is important and leads to a higher Jeans mass, which results in fewer and more massive stars \citep[][see also Sec.~\ref{RPandIR}]{2007Krumholz,2010Peters,2015Rosdahl,2017Howard,Hennebelle2020b}.\citep{2009bate} show that radiative heating reduces small-scale fragmentation and shifts the IMF towards higher masses, which helps explain why the IMF appears to be universal.

\subsubsection{Protostellar Outflows}
Stellar outflows from both low- and high-mass protostars can regulate mass accretion during star formation. Observations show that these outflows can remove material and angular momentum from the protostellar environment, carve cavities in the infalling envelope, and disrupt accretion flows \citep[e.g.][]{2015maud,2016Kuiper}. By ejecting gas from the system, they reduce the reservoir available to the forming star and inject turbulence into the surrounding medium, and thereby lowering the SFE \citep{Kuiper2015,2018Kolligan,Rosen2020,2021Rohde,Commercon2022,Oliva2023}.  In our simulations, protostellar outflows are not yet included which might lead to an overestimate of the accretion rates and final stellar masses. Outflows from massive sink particles could also hinder mass accretion onto the MMS and might limit its mass \citep{2018Tanaka}.
\subsubsection{Stellar Winds}
Also, stellar winds might regulate the SFE. These winds can reach velocities of thousands of kilometres per second and carry mass and momentum into the circumstellar medium \citep[e.g.][]{2008Puls}. By removing material from accretion flows and clearing gas from the immediate vicinity of the star, stellar winds have the potential to reduce accretion onto very massive stars \citep{2022Rosen, 2014Krumholz}.
However, \cite{2018Haid} find that stellar winds dominate over ionizing radiation only under conditions where the gas is already heated to $10^4 \, \rm K$. Moreover, hydrodynamic simulations suggest that stellar winds can generate cavities and bubbles that further disrupt star-forming clouds on larger scales \citep{2015Geen, 2014Dale}.
On smaller scales, line-driven disc ablation might play a role in preventing mass accretion onto massive stars \citep{2019KeeKuiper}.

\subsection{Magnetic Fields}
In addition to the feedback mechanisms, magnetic fields are an important factor in star formation, influencing fragmentation and accretion processes \citep{2019Hull,2012Crutcher}. By providing additional support against gravitational collapse and regulating angular momentum transport, magnetic fields reduce fragmentation and lower accretion rates onto protostars. Several Studies show that magnetic fields can affect both, the final stellar masses and the overall SFE \citep[e.g.,][]{2019Krumholz,Rosen2020,Hennebelle2022,mignon-risse2021, 2014Myers, 2015Seifried}. Therefore, including magnetic fields in our simulation would most likely lower the accretion rates, stellar masses and SFE.

\citep{Kuiper2015} emphasize that the mass accretion period and thus the final mass in sink particles highly depend on the lifetime of the circumstellar accretion disc and the surrounding material, which can be influenced by the processes mentioned above.

\subsection{Resolution Limitations}
In our simulations, the run with a spatial resolution of $\sim 200 ~\rm au$ produces a more massive sink than the run with a resolution of $\sim 400 ~\rm au$. The higher resolution allows us to resolve higher gas densities, which in turn affects the propagation of ionizing radiation. Since the recombination rate scales with the square of the density, ionizing photons are absorbed more efficiently at higher densities, effectively trapping the radiation for longer and delaying the expansion of the UC-HII region. This delay gives the sink more time to accrete additional mass before feedback-driven disruption of the cloud halts further accretion.
In our simulations, each sink particle represents a single star. The fiducial spatial resolution of $\sim 400 ~\rm au$ is not high enough to resolve the internal structure of massive stars or very close companions. This means that some sink particles probably contain several unresolved stars. Treating them as single objects increases the effective accretion, so that gas that would in reality be divided among multiple stars instead accretes onto a single sink, raising its accretion rate and final mass \citep{2009bate,federrath2010}. At the same time, the artificially high mass of these sinks results in a stronger radiative feedback, which can heat and disperse surrounding gas more efficiently, halting accretion earlier and suppressing further fragmentation \citep{Krumholz_2009, 2010bkuiper,2016rosen}.
These two effects act in opposite directions: unresolved multiplicity tends to boost accretion, while stronger feedback tends to limit it. In our simulations, the enhanced accretion dominates during the early phases, leading to more massive stars. As the most massive sinks grow, their intensified feedback becomes increasingly effective at heating and dispersing the surrounding gas, which can slow or shut down further accretion and reduce subsequent star formation \citep{2010Peters,2013myers}.
Our simulations resolve turbulent fragmentation, but lacks additional disc fragmentation, which might result in too massive stars, an overestimation of stellar feedback and a lack of lower-mass companion stars.

When simulations on a larger scale are performed, sink particles usually represent several stars or a star cluster. In this case, subgrid models that sample the stellar IMF are used \citep{Rathjen2023,KimKim2023}.
Using a coarser resolution than ours ($\sim 7000 ~\rm au$) \cite{2021Cournoyer} find that including a massive star binary fraction in clusters may have an impact on the dynamics of the forming star cluster. However, from observations \cite{2021Enokiya} sees a correlation between the number of massive stars in a region and the corresponding column density. In our simulation 11 massive sink particles are formed at a column density of around $1 \, \mathrm{g cm}^{-2}$ which is in agreement with \cite{2021Enokiya}.

\subsection{Massive Star Formation Scenarios}
There are several theories about how a massive star forms and how it accretes mass. The theory of a core collapse scenario suggests that a star is formed from a single clump that already contains most of the final stellar mass of the forming star \citep{mckeetan2003}. Less than half of the clump mass is supposed to end up in one massive star. This cannot be seen in our simulations, as our sink particles do not have an isolated mass reservoir and still accrete mass from their surroundings while moving towards the centre of the simulation domain.
However, the competitive accretion model \citep{2001Bonnell} says that the gas which is eventually accreted by a forming massive star is initially not necessarily gravitationally bound to the star. The formed stars share a mass reservoir and compete to accrete. This scenario can be found in our simulations. 
In addition, \cite{Padoan2020} propose the internal inflow model, where most of the final mass of a star reaches the star due to a random velocity field from larger scales and not due to its gravitational potential. Although they simulated a $(250 \, \mathrm{pc})^3$ box, this can also be seen on smaller scales in our simulation. Dense gas clumps move towards the most massive sink, where they are accreted. However, the clump motion can be caused by both turbulence and gravity; in our simulations, it is not clear that turbulence dominates the clump motion.

\section{Conclusions}
\label{sec:Conclusion}
This work studies (massive) star formation in isolated collapsing turbulent cores with an initial mass of $1000 \, \rm M_\odot$ and a radius of $1~ \mathrm{pc}$ with the hydodynamic AMR code FLASH.
A novel scheme to treat the radiative transfer of ionizing radiation and radiation pressure from ionizing radiation as well as heating by non-ionizing radiation is included.
The temperatures of gas, dust, and radiation are calculated self-consistently. 
We investigate the collapse, star formation, and evolution of the cores until UC-HII regions are established and star formation is mostly complete.

We analyze in detail the evolution of the \textsc{Fiducial} run, a core with solar metallicity, a moderately steep density profile, and a virial parameter of 0.6. In total 19 sink particles are formed, while 10 become massive.
Initially, the dust temperature follows the gas temperature morphology because of gas-dust coupling. Once sink particles form, radiative feedback heats the dust and the dust temperature primarily governed by the radiation temperature.
Due to turbulence, the accretion rates of sink particles are highly variable. The combination of the local and global collapse leads to the formation of the most massive sink, the so-called MMS.
The MMS forms near the centre and is only the 4$^{\rm th}$ sink particle that forms and dominates the cluster due to its mass. 
The density around the MMS is typically so high that ionizing radiation is trapped, while mass accretion is still ongoing. Ionizing feedback from other massive sinks, located in less dense environments, launches an expanding HII region, which finally disrupts the central hub around the MMS. Then, the trapped ionizing feedback from the MMS escapes, further powers the HII region, and finally disrupts the parental cloud.

We compare the feedback mechanism of ionizing radiation and radiation pressure. Ionizing radiation sets an upper limit on the mass accretion. Without ionizing feedback, sink particles do not stop accreting mass during the simulated time period. We find that the momentum, which is transferred to the gas by direct radiation pressure from ionizing radiation, supports the growth of the HII region. 
Radiative heating from non-ionizing radiation is essential, as it suppresses fragmentation while the core warms up.

Our resolution study shows that increasing resolution leads to an increasing number of (lower-mass) sink particles. At the same time, sink particles are formed at higher densities, thus ionizing feedback of the MMS is trapped for a longer time period, and thus the MMS has more time to accrete mass. As a result, the total SFE increases with resolution.

Finally, we investigate the impact of different initial conditions. All simulations are dominated by one very massive sink particle, which is most likely not the sink particle that forms first. 
The flatter the density profile, the slower the collapse. However, the final number of sink particles as well as the SFE increase for flatter core profiles. With a low virial parameter, gravity is more dominant than kinetic motions. This leads to a faster core collapse that produces fewer but very massive sink particles which have a high ionizing photon emissivity leading to a decrease in the overall SFE.
With an initially high virial parameter, turbulence is more dominant and the core collapse is time delayed. More substructures are produced, resulting in a slightly increased number of sink particles that have a lower average mass, resulting in a higher SFE.
Lower metallicity leads to reduced dust cooling, resulting in higher average temperatures during the initial collapse phase and in the suppression of fragmentation. Fewer but massive sink particles are formed, while in total the SFE decreases. A core with a higher metallicity can more efficiently cool the dust, resulting in more fragmentation and a higher SFE.
However, the SFE is very sensitive to the environment of the most massive stars. The longer massive stars are embedded and the formation of an UC-HII region is suppressed, the higher the overall SFE. 
Therefore, in our simulation a faster core collapse does not necessarily result in a higher SFE, as feedback plays a key role in determining the overall SFE.

The advanced and more realistic temperature calculation in our simulations provides an ideal test bed to compare with observations of high-mass cores such as observed in the ALMAGAL project. The comparison with observations will be presented in a follow-up paper.

\section*{Acknowledgements}
We thank the anonymous referee for their valuable comments on our manuscript.
We gratefully acknowledge the Gauss Centre for Supercomputing e.V. (www.gauss-centre.eu) for funding this project by providing computing time through the John von Neumann Institute for Computing (NIC) on the GCS Supercomputer JUWELS at Jülich Supercomputing Centre (JSC).
The software used in this work was in part developed by the DOE NNSA-ASC OASCR Flash Center at the University of Chicago.
BZ and SW gratefully acknowledge funding via the Collaborative Research Center 1601 (SFB 1601, sub-project A5) funded by the German Science Foundation (DFG). SDC is supported by the Ministry of Science and Technology (MoST) in Taiwan through grant MoST 108-2112-M-001-004-MY2. RW acknowledges support by project RVO:67985815. 

\section*{Data Availability}
The data underlying this article will be shared on reasonable request
to the corresponding author.



\bibliographystyle{mnras}
\bibliography{example} 




\appendix

\section{Additional Figures}
\begin{figure*}
	\includegraphics[width=1.0\textwidth,trim=0.0cm 0.0cm 0.0cm 0.0cm]{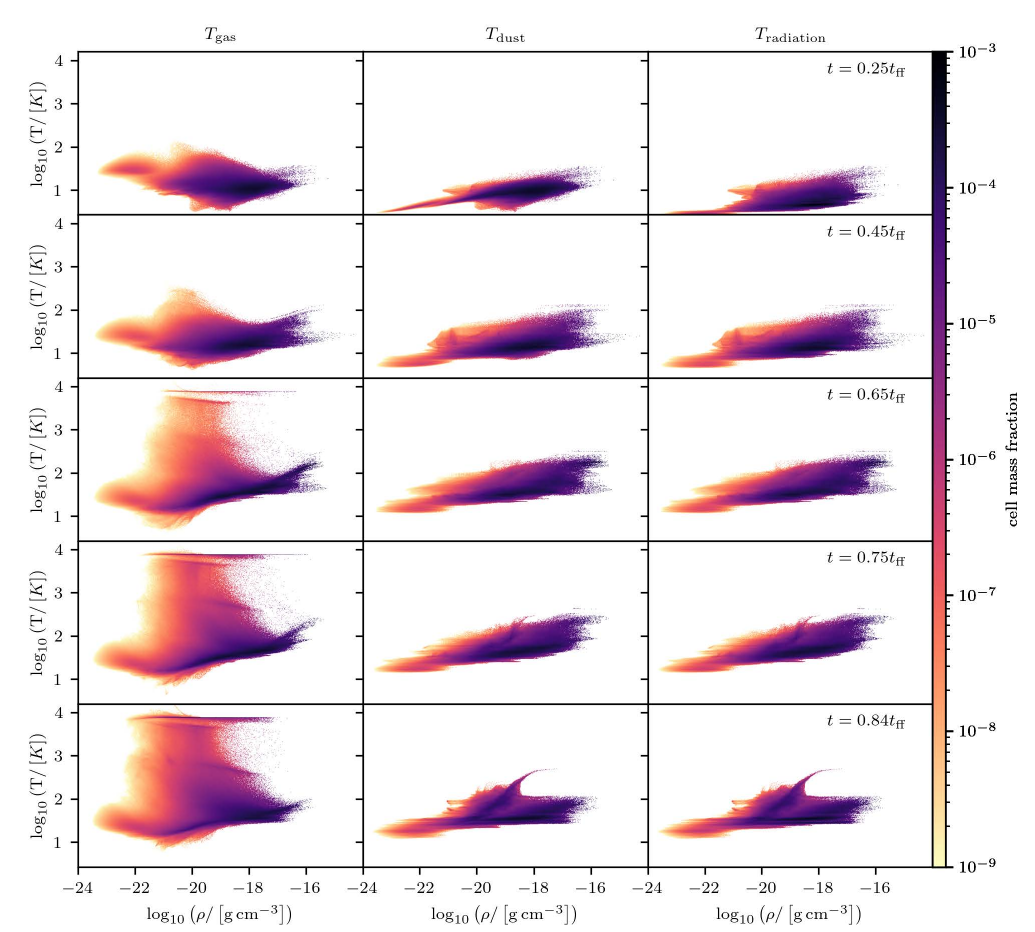}
    \caption{Phase diagram of the temperature evolution (from top to bottom) for gas, dust and radiation temperature (left to right) in the \textsc{Fiducial} run. At the beginning the radiation temperature is colder than the dust and gas temperature. After $\sim 0.4 \, t_\mathrm{ff}$ the dust and radiation temperature behave similar. The gas temperature rises (up to $10^4$ K) as soon as sink particles become massive (after $\sim 0.4 \, t_\mathrm{ff}$). As a result, the dust and radiation temperature become warmer. The very dense region ($10^{-16}$ g cm$^{-3}$) at $0.70 \, t_\mathrm{ff}$ shows the region around the MMS, before ionizing feedback disperses the dense central hub. } 
    \label{fig:fiducialrunTemp}
\end{figure*}

We show a phase diagram of the temperature evolution for gas, dust and radiation temperature in the \textsc{Fiducial} run (see Fig. \ref{fig:fiducialrunTemp}). The gas temperature is initially warmer than the dust and radiation temperature. However, after $\sim 0.2 \, t_\mathrm{ff}$, most of the gas and dust are in equilibrium. Later, the dust temperature follows the radiation temperature. When the sinks become massive, ionizing radiation is emitted and the gas temperature increases to up to $10^4$ K in regions with a gas density between $10^{-22}$ g cm$^{-3}$ and $10^{-17}$ g cm$^{-3}$. At the same time, the dust and radiation temperatures warm up. At $0.70 \, t_\mathrm{ff}$ the central hub around the MMS becomes very dense ($\sim10^{-16}$ g cm$^{-3}$) and warms before it is dispersed by ionizing feedback.\\

\begin{figure*}
	\includegraphics[width=0.9\textwidth]{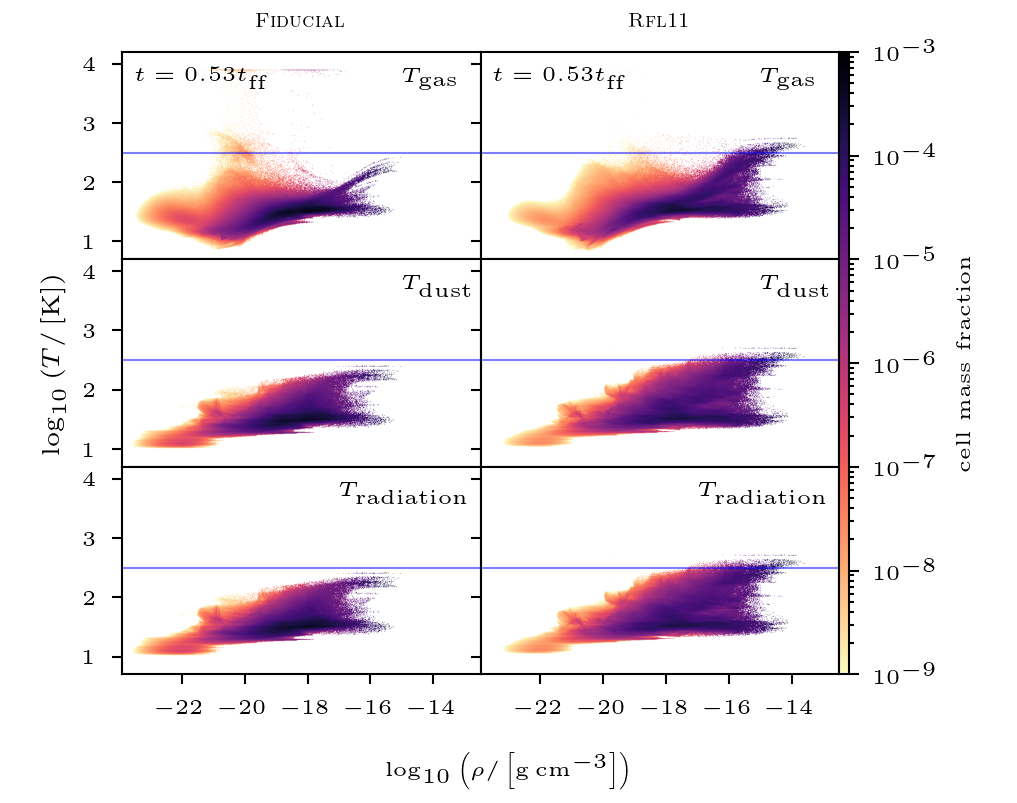}
    \caption{Phase diagram of the temperature for gas, dust and radiation temperature (top to bottom) for the runs \textsc{Fiducial} (left panels) and \textsc{Rfl11} (right panels). Blue lines help for better comparison. A higher resolution (run \textsc{Rfl11}) resolves higher densities and dense cells can be heated up to higher temperatures compared to the \textsc{Fiducial} run.}
    \label{Fig.resolutionT}
\end{figure*}

Fig. \ref{Fig.resolutionT} shows a temperature phase diagram of gas, dust, and radiation of two runs in the resolution study (run \textsc{Fiducial} and \textsc{Rfl11}) at $0.53 \, t_\mathrm{ff}$. Using a higher resolution (run \textsc{Rfl11}) leads to a higher density threshold for sink particles (see also section \ref{Sec:ResolutionStudy}) and therefore higher densities can be resolved. These higher densities reach higher gas, dust, and radiation temperatures compared to the \textsc{Fiducial} run even before ionizing radiation sets in.\\

\begin{figure}
	\includegraphics[width=1.0\columnwidth]{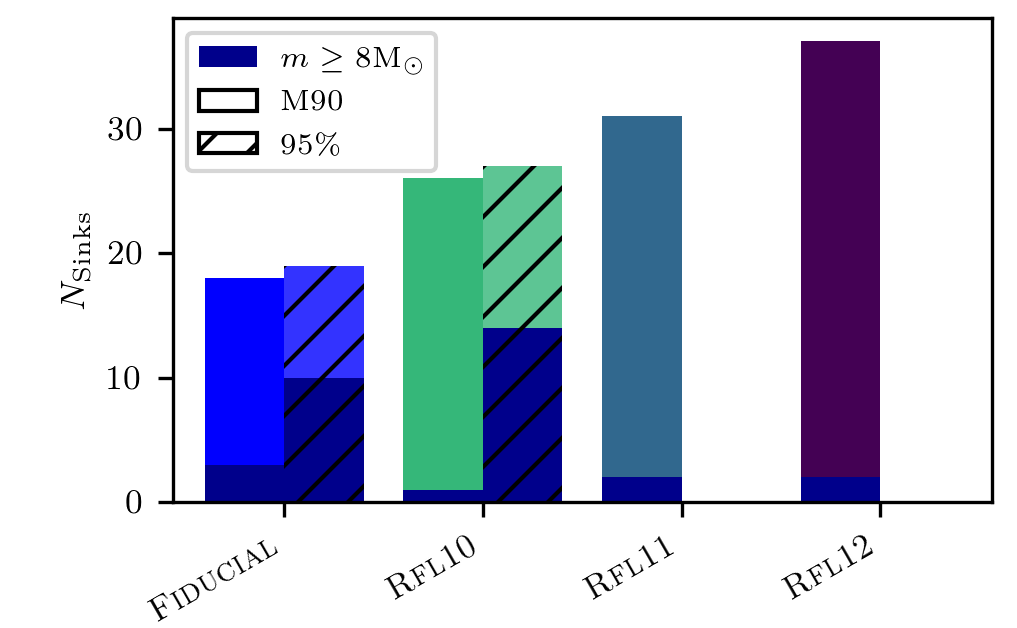}
    \caption{ Number of sink particles at two different stages, when $90 \, \mathrm{M}_\odot$ and $95 \%$ of the final mass has been accreted, shown by the non hatched and hatched region, respectively. Runs \textsc{RFL11} and \textsc{RFL12} never reach this stage due to computational costs. The dark blue area shows the number of high-mass stars. The total number of low- and high-mass sink particles increases with higher resolution.}
    \label{Fig.resolutionhist}
\end{figure}

Fig. \ref{Fig.resolutionhist} shows the number of sink particles for two different stages. First, when $90 \, \mathrm{M}_\odot$ are in sink particles and second, when $90 \%$ of the final mass has been accreted. Due to computational costs, runs \textsc{RFL11} and \textsc{RFL12} were stopped before they reached the second time. However, already when $90 \, \mathrm{M}_\odot$ have been accreted onto sinks we can see that an increasing resolution leads to the formation of more sink particles. Stage two ($95\%$) indicates that, especially, the number of low-mass sinks increases with higher resolution. Nevertheless, comparing both stages shows that when $90 \, \mathrm{M}_\odot$ are accreted onto sink particles the fragmentation process is almost complete and only one additional sink particle is formed in both, runs \textsc{Fiducial} and \textsc{RFL10}.\\

\begin{figure}
     \includegraphics[width=\columnwidth]{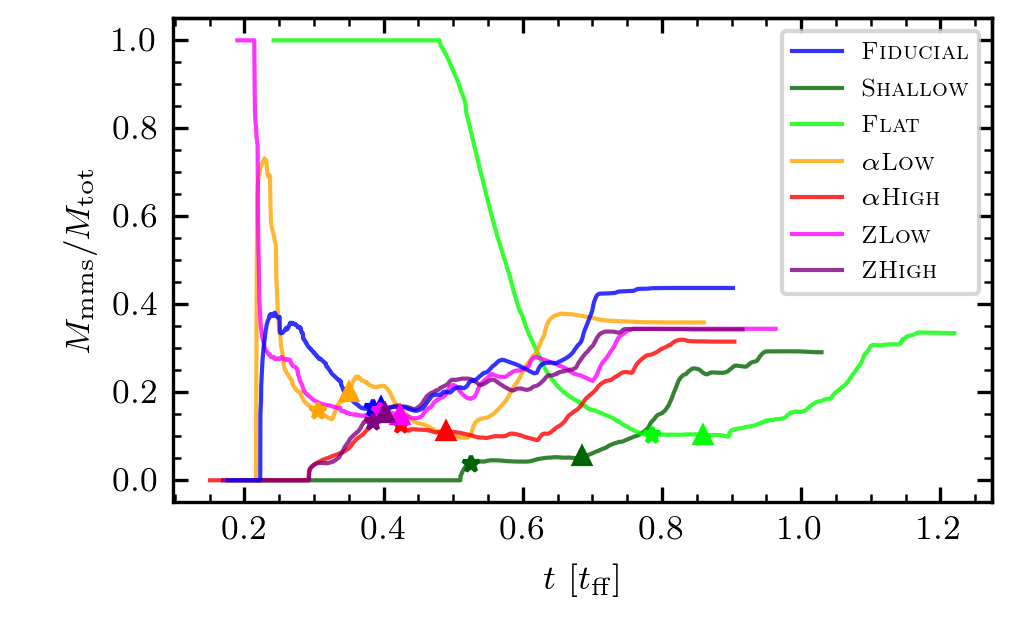}
     \caption{Fraction of the accreted mass onto the MMS normalized by the total accreted mass. The starts indicates when the first sink particle in the simulation reaches 8 M$_\odot$ and the triangles show when the MMS becomes massive.
     Only for two runs (\textsc{ZLow} and \textsc{Flat}) the first sink particle evolve into the most massive one. For every run the MMS is not the first sink which becomes massive.}
     \label{Fig.MassratioMMSparameter}
\end{figure}

In Fig. \ref{Fig.MassratioMMSparameter} we show the fraction of mass that has been accreted onto the MMS, normalized by the total mass accreted by all sink particles for the runs of the parameter study. The fist sink particles in each run do not necessarily evolve into the MMS. Only in runs \textsc{ZLow} and \textsc{Flat} the fist sink particle also evolves into the MMS. The time when the first sink particle in each simulation becomes massive is marked with a star. Triangles indicate when the MMS reaches $8 ~\rm M_\odot$. In all runs the MMS is overall not the first sink particle that becomes massive. \\
 
\begin{figure}
     \includegraphics[width=\columnwidth]{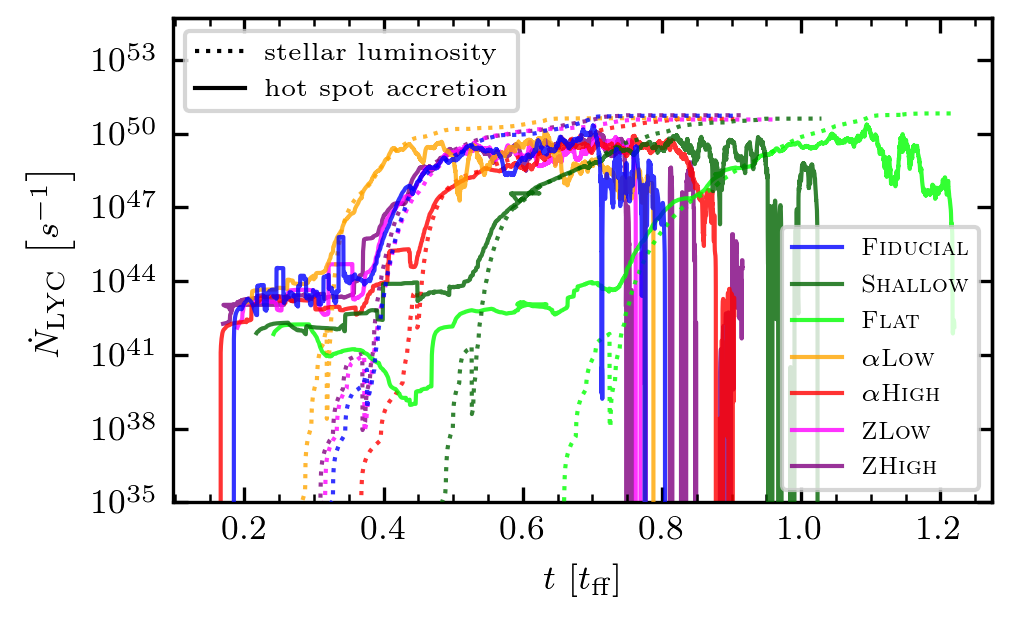}
     \caption{Total ionizing photon rate (smoothed) from hot spot accretion (solid lines) and stellar luminosity (dotted lines) for the runs in the parameter study. 
     In run \textsc{$\alpha$Low} the ionizing photon rate increases faster compared to the fiducial run, while it is time delayed for run \textsc{$\alpha$High} and runs with shallower density profiles (\textsc{Shallow} and  \textsc{Flat}). Later, the ionizing photon rate from stellar luminosity is comparable for all runs.}
     \label{Fig.parameter_total_LYC}
\end{figure}

Fig. \ref{Fig.parameter_total_LYC} shows the total ionizing photon rate for the runs in the parameter study. The ionizing photon rate highly depends on the mass growth of the sink particles (see Fig. \ref{fig:parameter}. High mass accretion rates and an early formation of massive sink particles generally lead to a rapid increase of the ionizing photon rate. Therefore, the ionizing photon rate increases faster for the run \textsc{$\alpha$Low} compared to the \textsc{Fiducial} run while the time-delayed evolution of the runs \textsc{$\alpha$High}, \textsc{Shallow}, and  \textsc{Flat} results in a time-delayed rise of the total ionizing photon rate. 
Although the time evolution of the ionizing photon rate varies, it eventually converges across all runs as the stellar luminosity becomes the dominant factor.

\bsp	
\label{lastpage}
\end{document}